\documentclass[conference]{IEEEtran}
\usepackage{amsmath}
\usepackage{multirow}
\usepackage[pdftex]{graphicx}
\usepackage[font=small]{subcaption}
\usepackage{subfig}

\begin{document}
\title{A Comparison of Methods for Cascade Prediction}
\author{\IEEEauthorblockN{Ruocheng Guo, Paulo Shakarian}
\IEEEauthorblockA{
Arizona State University\\
Tempe, AZ\\
Email: \{rguosni,shak\}@asu.edu}
}

\maketitle

\begin{abstract}
\label{abs}
Information cascades exist in a wide variety of platforms on Internet.
A very important real-world problem is to identify which information cascades can ``go viral''.
A system addressing this problem can be used in a variety of applications including public health, marketing and counter-terrorism.
As a cascade can be considered as compound of the social network and the time series. 
However, in related literature where methods for solving the cascade prediction problem were proposed, the experimental settings were often limited to only a single metric for a specific problem formulation. 
Moreover, little attention was paid to the run time of those methods.
In this paper, we first formulate the cascade prediction problem as both classification and regression. 
Then we compare three categories of cascade prediction methods: centrality based, feature based and point process based.
We carry out the comparison through evaluation of the methods by both accuracy metrics and run time.
The results show that feature based methods can outperform others in terms of prediction accuracy but suffer from heavy overhead especially for large datasets.
While point process based methods can also run into issue of long run time when the model can not well adapt to the data.
This paper seeks to address issues in order to allow developers of systems for social network analysis to select the most appropriate method for predicting viral information cascades.
\end{abstract}

\section{Introduction}
Identifying when a piece of information goes ``viral'' in social media is an important problem in social network analysis.
This is often referred to as ``cascade prediction''.
Recently, the cascade prediction problem attracted considerable attention from researchers from communities of machine learning, data mining and statistics.
Researchers attempted to predict the final size of information cascades based on approaches inspired by knowledge in various areas.
Pei et al.~\cite{pei2014searching} measured influence of the root node by k-shell number and related heuristics.
Weng et al.~\cite{weng2014predicting} and Guo et al.~\cite{guo2015toward} uitilized features describing both structural and temporal properties of early-stage cascades.
The work described in~\cite{zhao2015seismic} and~\cite{shen2014modeling} modelled cascades by one-dimensional point process.
 However, in this line of research, the experimental settings varied from paper to paper.
Furthermore, as the cascade prediction problem can be treated as either classification or regression, most of previous work only dealt with one or the other and using just a single evaluation metric.
With deployment of a counter-extremism messaging system (i.e. an enhanced version of~\cite{kim2013lookingglass}) as one of the primary goals in our group, cascade prediction can play a crucial role in detection of early-stage extremism message that is potential to go viral on social network sites.
Other applications include the spread of information following a disaster, promotion of health behaviors and applications to marketing.
Therefore, it is important to understand how well the existing methods stemming from different research area could perform in near real-world experimental settings.
An ideal cascade prediction method for counter-extremism messaging system should provide acceptable accuracy with ability to make near real-time prediction.

In this paper, we compare performance of a variety of cascade prediction methods originating from different research areas as both classification and regression problems with multiple evaluation metrics.
We also measure the run time of the tasks required by the methods to complete cascade prediction -- another key deployment concern not explored in most research.

In this paper, the main contribution can be summarized as:
\begin{itemize}
	\item We compare cascade prediction methods in three categories: centrality based, feature based and point process based, therefore providing comparison between methods orginating from different research areas.
	\item The cascade prediction problem is considered from both the aspect of regression and classification. we also conduct a comprehensive comparison between methods by various evaluation metrics.
	\item We also compare the run time of tasks needed for the cascade prediction methods are also measured in a task by task style.
	
\end{itemize}
The rest of this paper is organized as the follows:
In Section~\ref{sec:tech}, definitions relevant to the methods considered in this paper are introduced along with a formal problem statement of cascade prediction.
Section~\ref{sec:method} summarizes the mechanism of the three categories of cascade prediction methods.
Section~\ref{sec:setup} and~\ref{sec:result} presents the setup of experiments and performance of each method in terms of both accuracy and run time.
Section~\ref{sec:related} reviews related work.
At last, Section~\ref{sec:conclusion} concludes the paper and discusses the main issues of these methods.

\section{Technical Preliminaries}
In this section, related concepts for the three categories of methods are defined. Then we formulate the cascade prediction problem as regression and classification respectively. 
\label{sec:tech}
\subsection{Definitions}
\noindent\textbf{Network and Cascade:}
The social network is a directed graph $G=(V,E)$ where each node $v \in V$ represents a user and each edge $e_{ij} = (v_i,v_j)$ denotes that user $v_i$ is followed by user $v_j$.
Identified by the original message or the corresponding hashtag, a cascade is a time-variant subgraph of the social network $d(t) = (V(t),E(t))$. Each node $v \in V(t)$ denotes a user reposted the original message of cascade $d(t)$ (for the Aminer dataset \cite{zhang2013social}) or a user posted the hashtag defining cascade $d(t)$ (for the Twitter dataset \cite{weng2014predicting}) within time $t$. The time variable $t$ denotes number of time units since the microblog including the original message or the hashtag. For each node $v \in V(t)$ we record their adoption time of cascade $d(t)$ as $t_v$. For $v \in V(t)$, $t_v \le t$ while for $v \not \in V(t)$ we define $t_v = \infty$. Thus we can get an ascendingly sorted vector $\mathbf{t_v}(t)$ including all $t_v \le t$ for each cascade, which plays an important role in both feature based methods and point process based methods for cascade prediction. The $kth$ element of $\mathbf{t_v}(t)$ can be denoted as $\mathbf{t_v}(t)[k]$. For convenience, we use $t_{end}$ to denote the time when the last adoption of a cascade happened.

Besides the cascade $d(t) = (V(t),E(t))$, the neighborhood of $V(t)$ also can provide information about the potential of the cascade. Here we define the out-neighborhood reachable by any node in $V(t)$ in step $i$ as ith surface $F_i(t)$.
To show how 'fresh' the cascade is for a node $v \in F_i(t)$, we define a function $f_{\Delta t} : v \rightarrow \Delta t$ that maps such a node to the number of time units since $v$ become a member of first surface to current time $t$.
As time makes a big difference in social influence and diffusion, we divide the first surface $F_1(t)$ into two sets of nodes depends on $f_{\Delta t}(v)$ for all $v \in F_1(t)$. With a selected threshold $t_{\lambda}$. The first set named as \textbf{frontiers} includes all nodes $v \in F_1(t)$ such that $f_{\Delta t}(v) \le t_\lambda$ and the other set \textbf{non-adopters} consists the other nodes $v \in F_1(t)$ with $f_{\Delta t}(v) > t_\lambda$.
In this paper, $|x|$ denotes absolute value of scaler $x$ and $\left\vert\mathbf{x}\right\vert$ denotes cardinality of set $\mathbf{x}$.

\noindent\textbf{Communities:}
We can treat a community partition of a social network as a function $f_C$ : $V \rightarrow C$ which maps a set of nodes $V$ to a set of communities $C$.
With this function, given a cascade $d(t) = (V(t),E(t))$, it enables us to describe the distribution of nodes over communities by features such as $ |f_C(V)| $, the number of communities among set $V$.

\noindent\textbf{Point Process:}
Each adoption in a cascade can be represented as an event from the aspect of point process as in~\cite{zhao2015seismic}.
Thus, for cascade prediction, we can use $\mathbf{t_{v}}(t-\Delta t)$ to describe the history of a point process strictly before t.
The core of a point process is the conditional density function $\lambda(t)$. Conditioned on $\mathbf{t_{v}}(t-\Delta t)$, the conditional density is the limit of expected number of adoptions would happen in time interval $\left[t,t+\Delta t\right]$ by taking $\Delta t \rightarrow 0^+$:
\begin{equation}
	\lambda(t) = \lim_{\Delta t \rightarrow 0^+} \mathrm{E}\left\{\left\vert V(t+\Delta t)\right\vert - \left\vert V(t) \right\vert\right\}
	\label{eq1}
\end{equation}

Given the density function $\lambda(t)$ and target prediction time $t'$, the predicted cascade size can be computed by:
\begin{equation}
	| \hat{V(t')} | = \left\vert V(t) \right\vert + \int_{t}^{t'} \lambda(\tau) d\tau
	\label{eq2}
\end{equation}

\subsection{Problem Statement}
In this paper, we focus on comparison of different methods which can solve the cascade prediction problem.
This problem can be formulated as either a regression problem or a classification problem:

\noindent\textbf{Regression Problem:} Given a early stage cascade $d(t) = (V(t),E(t))$ and the corresponding node attribute vector $\mathbf{t_v}(t)$ with constraint $\left\vert V(t)\right\vert = n$, the target is to predict the final size of the cascade $\left\vert V(t_{end})\right\vert$.

\noindent\textbf{Classification Problem:} A threshold $Thres$ is selected to label each cascade. For a given cascade if its $|V(t_{end})| \ge Thres$, we define it as a viral sample labeled as 1, otherwise, we label it as non-viral labeled as 0.
Then the problem is to classify a given early-stage cascade $d(t)$ to the viral class or the non-viral class.

\section{Methods}
\label{sec:method}
In this section we introduce several recently published methods for solving the cascade prediction problem.
Diffusion process in social network includes information of time series, network structure, sometimes with microblog content and node attributes, therefore, methods originated from knowledge in various research area like social network analysis, random point process and non-linear programming can be applied.
The methods can be categorized into: centrality based methods, feature based methods and point process based methods.
\subsection{Centrality Based Methods}
Previous work~\cite{pei2014searching} discovered that the k-shell value of a node is highly correlated to the average cascade size it initiates. In this paper, we also consider eigenvector centrality, out-degree and Pagerank of the root node of cascades to deal with the cascade prediction problem. We refer to centrality based approaches as method C in this paper.
\subsection{Feature Based Methods}
In this paper, we consider two recently proposed methods \cite{guo2015toward} and \cite{weng2014predicting} and call them method A and method B respectively for convenience.
The features computed by the two methods can be categorized into network features, community based features and temporal features.

Both of the feature based methods require to take advantage of community detection algorithms.
Given the social network, community detection algorithms such as \cite{blondel2008fast} and \cite{rosvall2008maps} can be applied to it and assign each node to one or multiple communities.
Based on the communities detected, features can be computed to numerically describe how the nodes that participate in a cascade are distributed over communities. Thus, we can quantitatively measure \textit{structural diversity} from~\cite{ugander2012structural} or \textit{influence locality} from~\cite{zhang2013social} as features.

\noindent \textbf{Network Features}: In method B proposed by \cite{weng2014predicting}, the authors consider several types of network features:
\begin{itemize}
	\item Neighborhood size, including first surface ($\left\vert F_1(V_t) \right\vert$) and second surface ($\left\vert F_2(V_t) \right\vert$).
	\item Path length, consisting average step distance and coefficient of variation of it, and diameter of the cascade. Step distance is the length of shortest path between two consecutive adopters $v_i$ and $v_{i+1}$.
\end{itemize}
Where coefficient of variation is defined as the ratio of the standard deviation to the mean.

\noindent \textbf{Community Based Features}: In both \cite{guo2015toward} and \cite{weng2014predicting}, community features are extracted and contribute to the predictive methods.
\begin{itemize}
	\item Community features for adopters, including the number of communities ($\left\vert f_C(V(t))\right\vert$), entropy and gini entropy.
	\item Community features for frontiers and non-adopters, including the number of communities ($\left\vert f_C(F_1(t))\right\vert$), entropy and gini entropy.
	\item The number of shared communties between any two groups of adopters, frontiers and non-adopters.
\end{itemize}

\noindent \textbf{Temporal Features}: In \cite{guo2015toward}, the authors computed average of $\mathbf{t_v}(t)$ while average step time and its corresponding coefficient of variation are calculated in \cite{weng2014predicting} as two features. 

\subsection{Point Process Based Methods}
To discover patterns in the temporal dynamics of cascades, authors of both~\cite{shen2014modeling} and~\cite{zhao2015seismic} both consider a cascade as an instance of one-dimensional point process in time space. They proposed novel density functions to characterize time series of cascades.
The two methods are quite similar, in terms of the formulation of conditional density function $\lambda(t)$. In both cases, $\lambda(t)$ consists of an element modeling the popularity of the cascade and another describing the probablity distribution of an adoption behavior over time. 

\noindent \textbf{The Reinforced Poisson Process (RPP) Method}: In~\cite{shen2014modeling}, the authors consider the density function for a cascade $d$ as a product of three elements:
\begin{equation}
	\lambda_d(t) = \alpha_d f_d(t;\theta_d)\left\vert V(t)\right\vert
	\label{eq3}
\end{equation}
For cascade $d$, $\alpha_d$ denotes the intrinsic attractiveness, $f_d(t;\theta_d)$ is defined as the relaxation function which models how likely an adoption would happen at time $t$ without considering $\alpha_d$ and $|V(t)|$.
For each cascade $d$, parameters $\alpha_d$ and $\theta_d$ are learned by maximization of the likelihood of $\mathbf{t_v}(t)$.
Thus, the predicted cascade size at time $t' > t$ can be computed by:
\begin{equation}
	| \hat{V}(t') | = \left\vert V(t) \right\vert + \int_{t}^{t'} \alpha_d f_d(\tau;\theta_d)\left\vert V(\tau) \right\vert d\tau
	\label{eq4}
\end{equation} 

\noindent \textbf{The SEISMIC Method}:
In~\cite{zhao2015seismic}, authors model the density function as a modified Hawkes Process made up of three elements: infectiousness $p_t$, node degree $n_i$ and human reaction time distribution $\phi(s)$:
\begin{equation}
	\lambda(t) = p_t \sum_{i=1}^{|V(t)|}n_i\phi(t-t_{vi})
	\label{eq5}
\end{equation}
Where $t_{vi} \in \mathbf{t_v}(t)$ is the time when each adoption happens.
Similar to $\alpha_d$ in the Reinforced Poisson Process model, $p_t$ is computed by maximization of the likelihood function: 
\begin{equation}
	p_t = \arg \max_{p_t} \prod_{i=0}^{|V(t)|-1} \lambda(t_{vi}) \exp^{-\int_{t_{vi}}^{t_{vi+1}} \lambda(\tau) d\tau }
\end{equation}
While the human reaction time distribution $\phi(s)$ is formulated as a piece-wise function consists of a constant piece and a power-law piece with parameter $c$ and $\theta$:
\begin{equation}
	\phi(s) = \begin{cases} 
	c & s \le s_0\\
	c(\frac{s}{s_0})^{-(1+\theta)} & s > s_0 \\
	\end{cases}
	\label{eq7}
\end{equation} 
As $\phi(s)$ is a probability distribution function, with the constraint $\int_{0}^{\infty} \phi(s) ds = 1$ and power-law decay factor $\theta$ estimated by training data, $c$ can be computed.
With the density function $\lambda(t)$, the predicted cascade size can be computed by equation~\eqref{eq2}.

\section{Experimental Setup}
\label{sec:setup}
For comprehensiveness, we evaluate the performance of each method by treating cascade prediction problem as both regression and classification problem. We only consider cascades that end up with at least 50 adopters.
Thus we can treat first 50 nodes of each cascade as its early stage.
In this section, an introduction of the datasets is followed by descriptions of setup of the classification and regression experiments.
All the experiments are carried out on an Intel(R) Xeon(R) CPU E5-2620 @
2.40 GHz machine with 256GB RAM running Windows 7.
All the methods are implemented in Python 2.7.
\subsection{Dataset Description}
The statistics of the two datasets used in this paper for evaluation of the cascade prediction methods are shown in Table~\ref{tbdataset}.

\noindent\textbf{Twitter Dataset:}
Twitter\footnote{https://twitter.com} is the most well-known microblog platform throughout the world. The dataset was used in~\cite{weng2014predicting}. This dataset includes a friendship network with undirected edges, cascades identified by hashtags and corresponding mentions and retweets.

\noindent\textbf{Weibo Dataset:}
Sina Weibo\footnote{https://weibo.com} is the largest Chinese microblog social network. The dataset was used in~\cite{zhang2013social}. It consists of a directed followership network and retweet cascades.

\begin{table}
	\caption{Dataset Statistics}
	\centering
	\renewcommand{\arraystretch}{1.3}
	\begin{tabular}{ | p{3cm}||c||c| } 
		\hline
		Property & Twitter Dataset & Weibo Dataset \\
		\hline
		Directed & undirected & directed \\
		Nodes & 595,460 & 1,787,443 \\ 
		Edges &  7,170,209 & 216,511,564 \\
		Number of communities & 24,513 & 2,802 \\
		Modularity & 0.7865 & 0.5581\\
		Average Out-degree & 47.94 & 231.3381\\
		Average Eigenvector Centrality & 0.001783 & 0.0186\\
		Average K-shell & 24.6032 & 52.3064\\
		Average Pagerank & $1.6794e^{-6}$ & $5.596e^{-7}$ \\
		Cascades ($\ge 50$ nodes) & 14,607 & 99,257\\
		\hline
	\end{tabular}
	\label{tbdataset}
\end{table}

\subsection{Regression}
For the regression problem, the $m \times 1$ ground truth vector $\mathbf{y}$ is made up of final size of each cascade ($|V(t_{end})|$), where $m$ is the number of cascade.
Each regression model is able to output a $m \times 1$ vector $\mathbf{\hat{y}}$. Thus each element $\hat{y}_i \in \mathbf{\hat{y}}$ is the predicted size of the ith cascade.
For point process models, with different prediction time, the predicted results can change. Thus, for each early-stage cascade, we set $t$ as the time when we observed the $50th$ adoption and prediction time as $\left\{2,4,6,8,10\right\} \times \mathbf{t_v}(t)[50]$.
To evaluate a method for the regression problem, the difference between its prediction results $\mathbf{\hat{y}}$ and the ground truth $\mathbf{y}$ can be described by various error functions.
In addition, $\mathbf{\hat{y}}_{top10\%}$ denotes the set of top 10\% cascades in prediction result while $\mathbf{y}_{top10\%}$ is the set top of 10\% cascades of ground truth.
In this paper we choose following metrics to compare the prediction made by different methods, as they are widely used in related literatures such as \cite{yu2015micro}, \cite{shen2014modeling}, \cite{gao2015modeling} and \cite{zhao2015seismic}: 
\begin{itemize}
	\item APE (average percentage error): $\frac{1}{m}\sum_{i=1}^m \frac{|\hat{y_i}-y_i|}{y_i}$
	\item RMSE (root mean square error): \[\sqrt{\frac{\sum_{i=1}^{m}(\hat{y_i}-y_i)^2}{m}}\] 
	\item RMLSE (root mean logrithm square error):\[\sqrt{\frac{\sum_{i=1}^{m}(\log \hat{ y_i}-\log y_i)^2}{m}}\]
	\item Top 10\% coverage: $\frac{10}{m}\left\vert \mathbf{\hat{y}}_{top10\%} \cap \mathbf{y}_{top10\%}\right\vert$
\end{itemize}

\subsection{Classification}
For classification, we apply three predetermined thresholds (50th, 75th and 90th percentiles) to final size of cascades to assign each of them a class label, which provides the $m \times 1$ ground truth vector $\mathbf{L} = \left\{l_0,...,l_{m-1}\right\}$ one for each threshold.
The cascades with size larger than threshold are labelled as viral class with $l_i = 1$.
Table~\ref{tbclf} shows the thresholds and counts of samples for both classes. Then the methods for solving the classification problem can output predicted label vector $\mathbf{\hat{L}}$. Comparing $\mathbf{L}$ with $\mathbf{\hat{L}}$ results in standard metrics: precision, recall and F1 score. To examine the effectivess of the methods, we focus on reporting the metrics on the minority class (viral) as it is more difficult to do good predictions for it than the other. 

Specially, for point process based mothods, as they are capable to predict the final cascade size without being trained with class labels (once parameters are determined and prediction times are selected), we carry out the evaluation on them in this way: prediction results (by setting different prediction times) are treated as features for each sample. As the time when each cascade stop growing is not easy to determine.

\subsection{Run time} 
We also take the run time of tasks into account for the cascade prediction methods. To understand how computationally expensive the methods are in terms of run time, it is necessary to analyze the procedure of them.
For centrality based methods, the prediction can be divided into three steps: computation of centrality, training and prediction.
Similarly, for feature based methods, computation of features, training and prediction are required to be done. In addition, preprocessing like community detection, computation of shortest path length are needed, which can be computationally expensive.
While point process based methods require little preprocessing. For each cascade, parameters are computed independently through MLE of the observed time vector $\mathbf{t_{v}}(t)$ and properties of the adopters $V(t)$. Then prediction is made by integral of density functions.
Thus, we consider the following processes one by one and then combine them to estimate the run time of a certain method.

\noindent\textbf{Proprecessing:}
There are three types of proprecessing considered: loading the graph, computation of centralities and community detection.

\noindent\textbf{Computation of Features:}
For feature based methods, we measure the run time of computation of the features , which takes the product of preprocessing as input.

\noindent\textbf{Training and Prediction:}
For centrality and feature based methods, the run time of training and prediction is measured for ten-folds.
For point process based methods, we measure the run time of parameter estimation and prediction for the whole batch of data.

\begin{table}
	\caption{Thresholds for Classification}
	\centering
	\renewcommand{\arraystretch}{1.3}
	\begin{tabular}{ | c||c||c||c | } 
		\hline
		Percentile & Threshold & Viral samples & Non-viral samples\\
		\hline
		\multicolumn{4}{|c|}{Twitter Dataset} \\ 
		\hline
		50\% & 106 & 7,303 & 7,304 \\ 
		75\% & 226 & 3,652 & 10,955 \\
		90\% & 587 & 1,461 & 13,146 \\
		\hline
		\multicolumn{4}{|c|}{Weibo Dataset}\\ 
		\hline
		50\% & 152 & 49,628 & 49,629 \\ 
		75\% & 325 & 24,814 & 74,443 \\
		90\% & 688 &  9,925 & 89,332 \\
		\hline
	\end{tabular}
	\label{tbclf}
\end{table}

\section{Experimental Results}
\label{sec:result}
\begin{figure}
	\centering
	\begin{subfigure}{.24\textwidth}
		\centering
		\includegraphics[height=2.5cm]{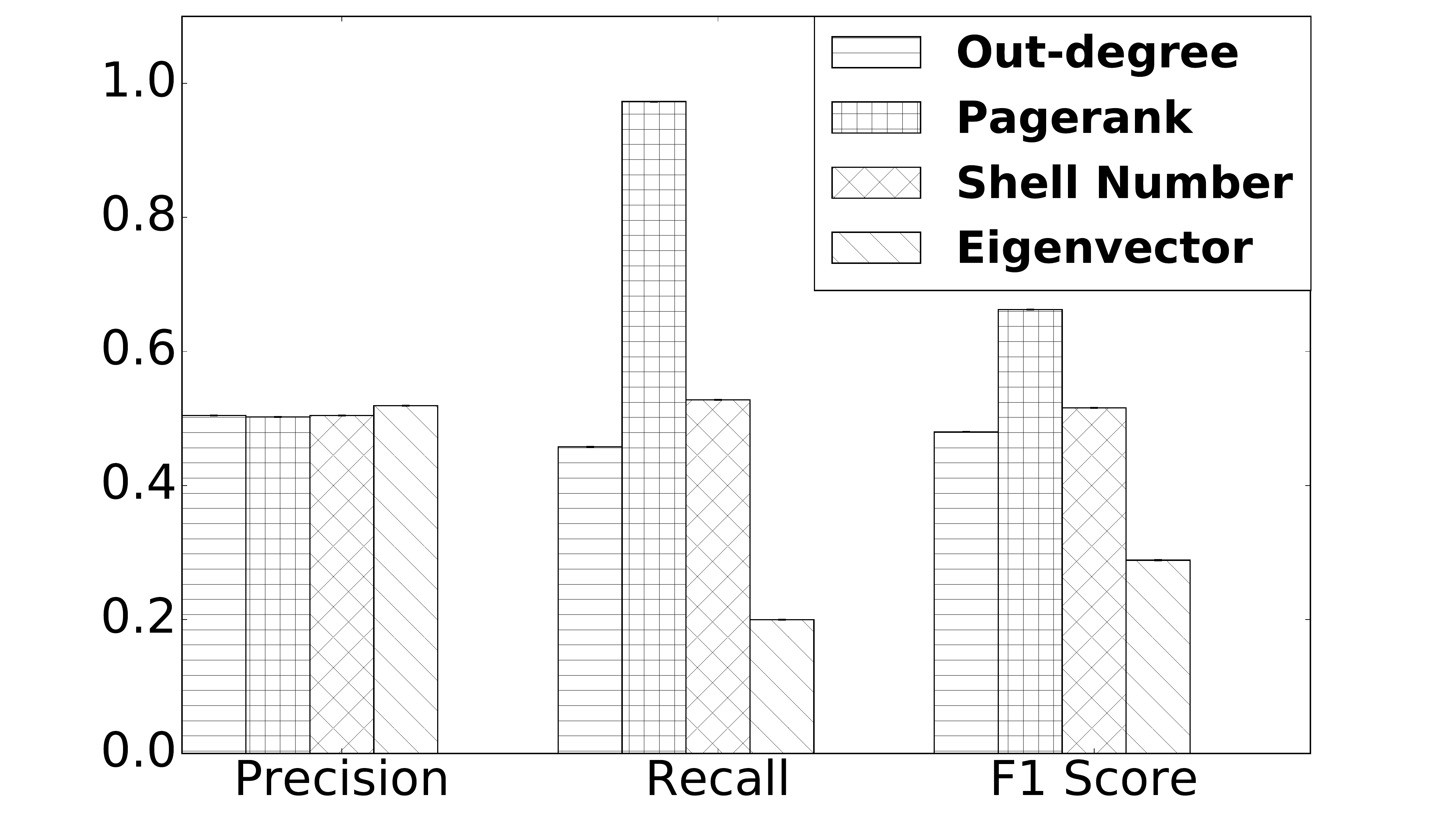}
		\caption{Twitter Dataset: 50th percentile}
		\label{fig:cent_clf_th0}
	\end{subfigure}%
	\hfill
	\begin{subfigure}{.24\textwidth}
		\centering
		\includegraphics[height=2.5cm]{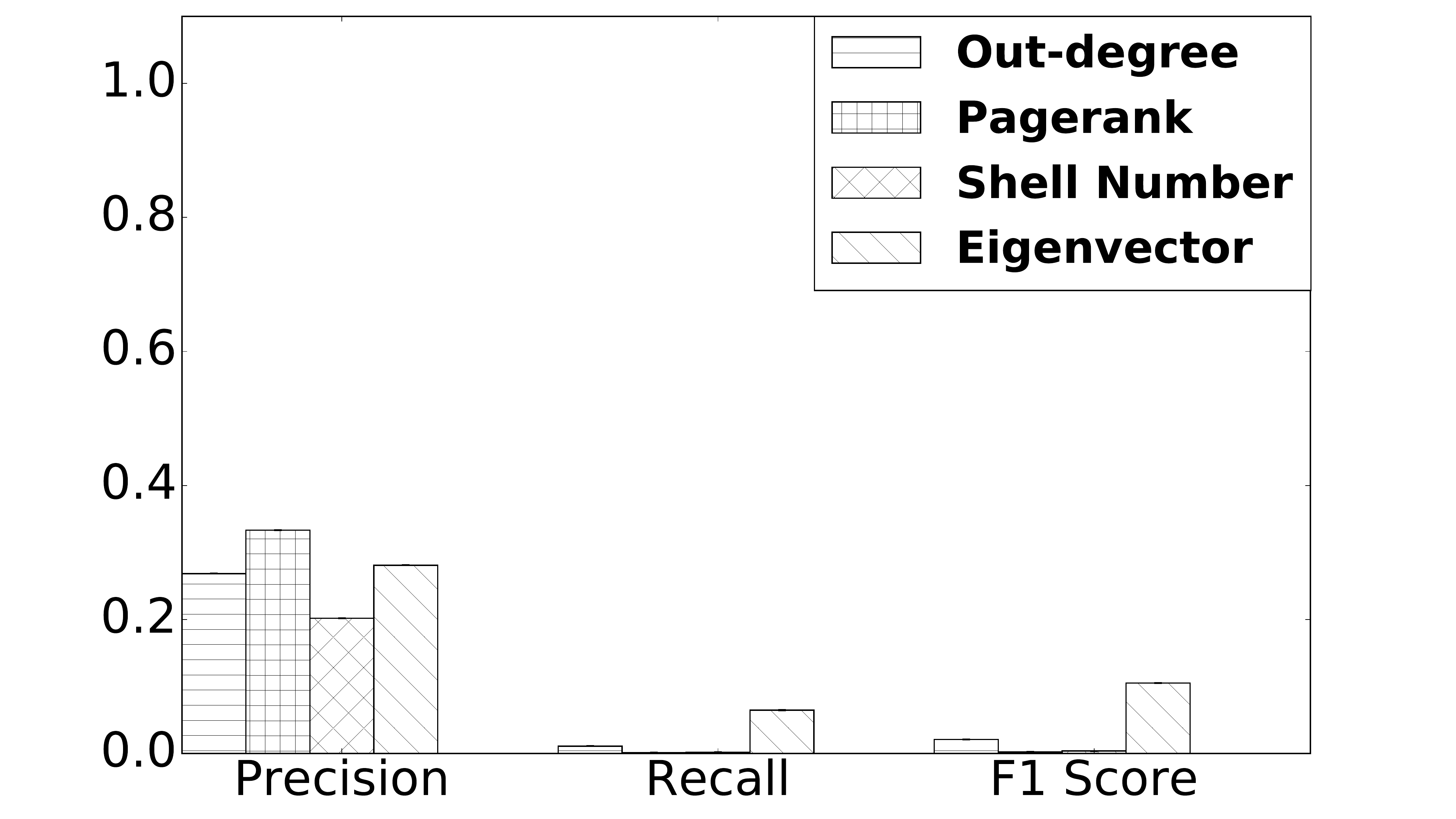}
		\caption{Twitter Dataset: 75th percentile}
		\label{fig:cent_clf_th1}
	\end{subfigure}
	\begin{subfigure}{.24\textwidth}
		\centering
		\includegraphics[height=2.5cm]{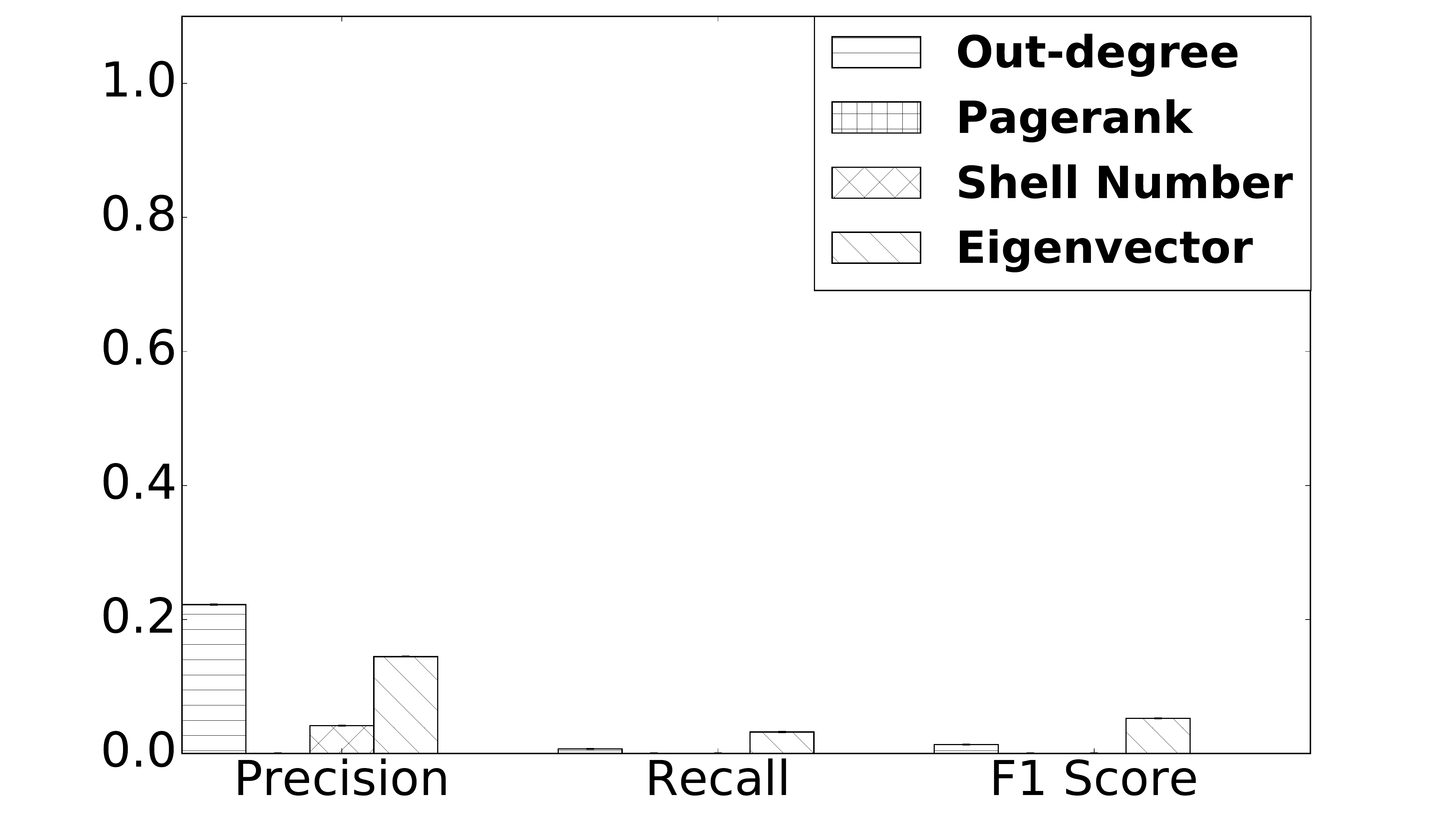}
		\caption{Twitter Dataset: 90th percentile}
		\label{fig:cent_clf_th2}
	\end{subfigure}
	\begin{subfigure}{.24\textwidth}
		\centering
		\includegraphics[height=2.5cm]{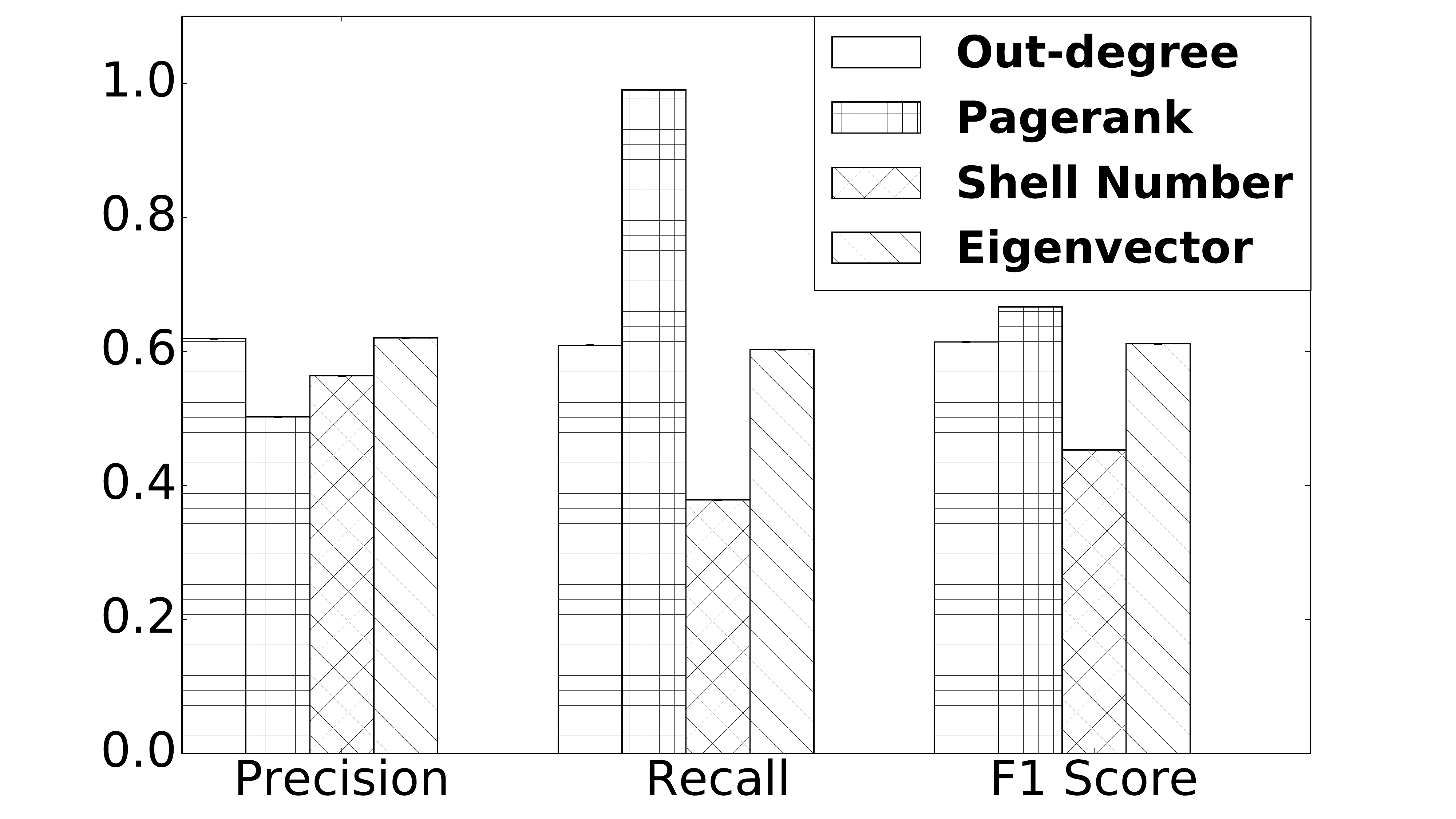}
		\caption{Weibo Dataset: 50th percentile}
		\label{fig:cent_clf_th0_w}
	\end{subfigure}%
	\hfill
	\begin{subfigure}{.24\textwidth}
		\centering
		\includegraphics[height=2.5cm]{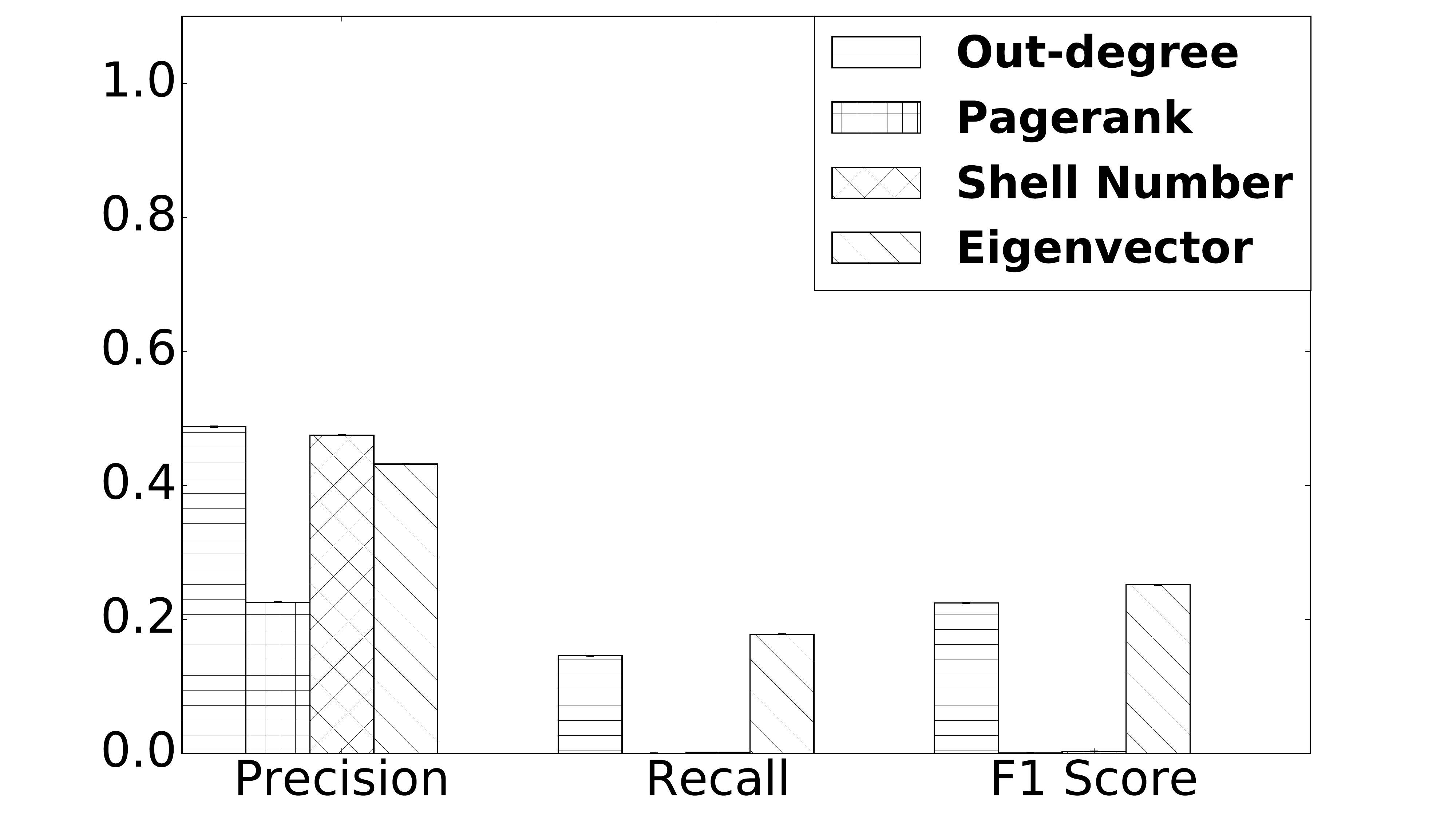}
		\caption{Weibo Dataset: 75th percentile}
		\label{fig:cent_clf_th1_w}
	\end{subfigure}
	\begin{subfigure}{.24\textwidth}
		\centering
		\includegraphics[height=2.5cm]{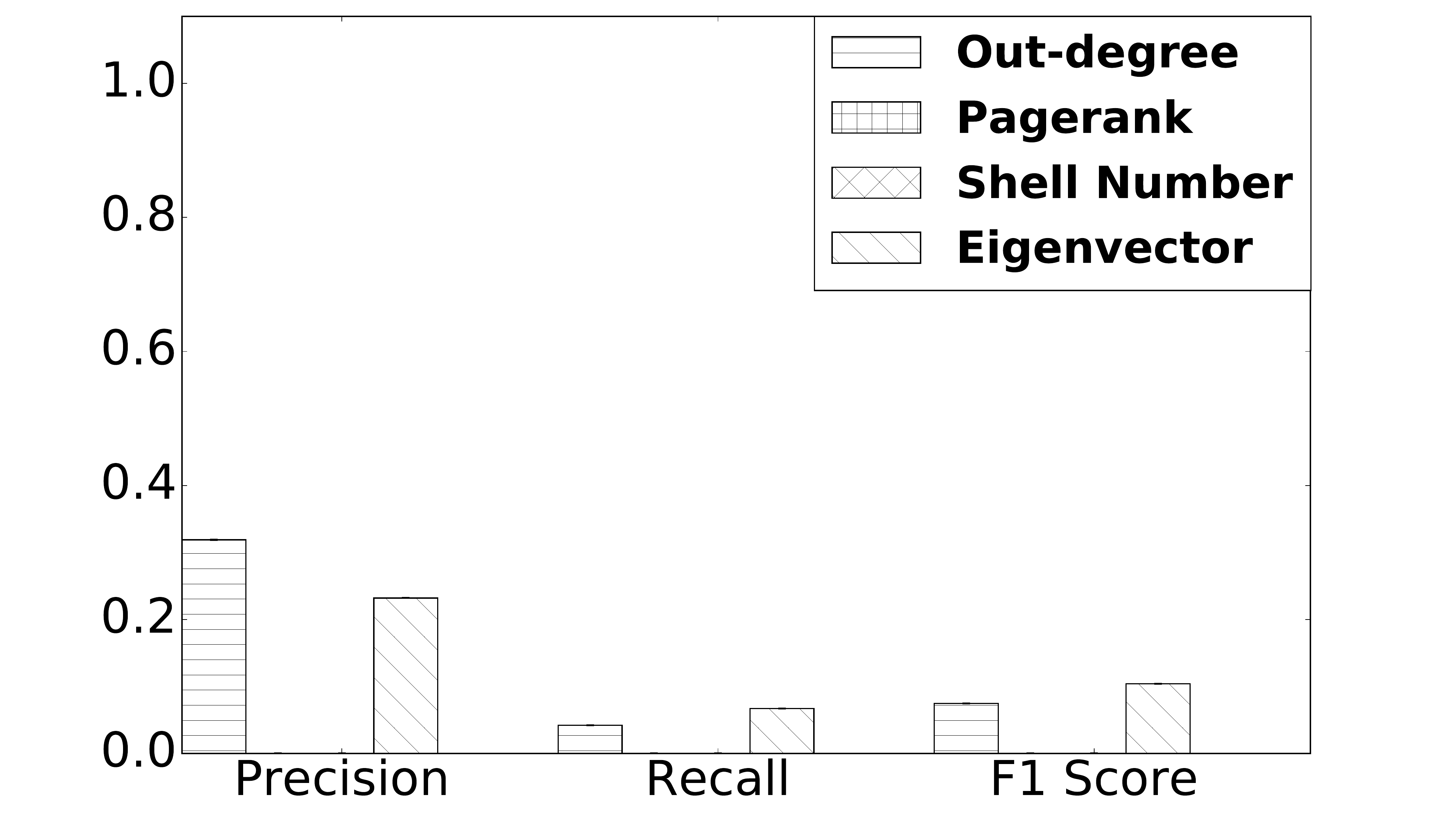}
		\caption{Weibo Dataset: 90th percentile}
		\label{fig:cent_clf_th2_w}
	\end{subfigure}
	\caption{Classification results of centrality based methods: error bar stands for one standard deviation.}
	\label{fig:cent_clf}
\end{figure}

In this section we show the experimental results including both accuracy of cascade prediction and the run time for each method.
For convenience, we call method of~\cite{guo2015toward}, ~\cite{weng2014predicting} and the centrality based method as method A, B and C respectively.
For method A, B and C, 10-fold cross-validation is applied.
For results where we compare these three methods,
we report only the best-performing centrality measure amongst out-degree, Pagerank, Shell number and eigenvector centrality as the method C for each dataset.
As shown in Fig.~\ref{fig:cent_clf}, eigenvector centrality outperforms others in the classification task when the two classes are imbalanced. Thus we take eigenvector centrality as the method C. The results for regression is not shown here for limited space as the difference between results produced by different centralities is trivial.
For the Reinforced Poisson Process (RPP) method~\cite{shen2014modeling}, as the parameter estimation task for each cascade is independent of others, the cross-validation is skipped and predictions are made by parameters learned from first 50 nodes of each cascade.
For the SEISMIC method~\cite{zhao2015seismic}, we also skip the 10-fold cross-validation. We set the cutoff time $s_0 = 30000(s)$ for the Twitter dataset and $s_0=300(s)$ for the Weibo dataset then fit the parameters $(\theta,c)$ for the human reaction time distribution function $\phi(s)$ with all samples in the dataset. While in the original paper~\cite{zhao2015seismic}, the authors set $\theta$ and $c$ just by 15 tweets they manually picked.
The power-law fitting is done as per~\cite{alstott2014powerlaw}, which returns $(\theta,c) = (0.440,1.018e^{-5})$ and $(0.282, 7.332e^{-4})$ for the Twitter dataset and Weibo dataset respectively.

\subsection{Regression}
For centrality based methods, we apply linear regression with least squared error.
We carry out the training and prediction with random forest regressor, SVR and linear regression model provided by~\cite{scikit-learn} for feature based methods. We only show the results produced by SVR as it outperformes others.
For the point process based mothods, we only report the best result among prediction time out of $\left\{2,4,6,8,10\right\} \times \mathbf{t_v}(t)[50]$.

\begin{figure}
	\centering
	\begin{subfigure}{.24\textwidth}
		\centering
		\includegraphics[height=2.5cm]{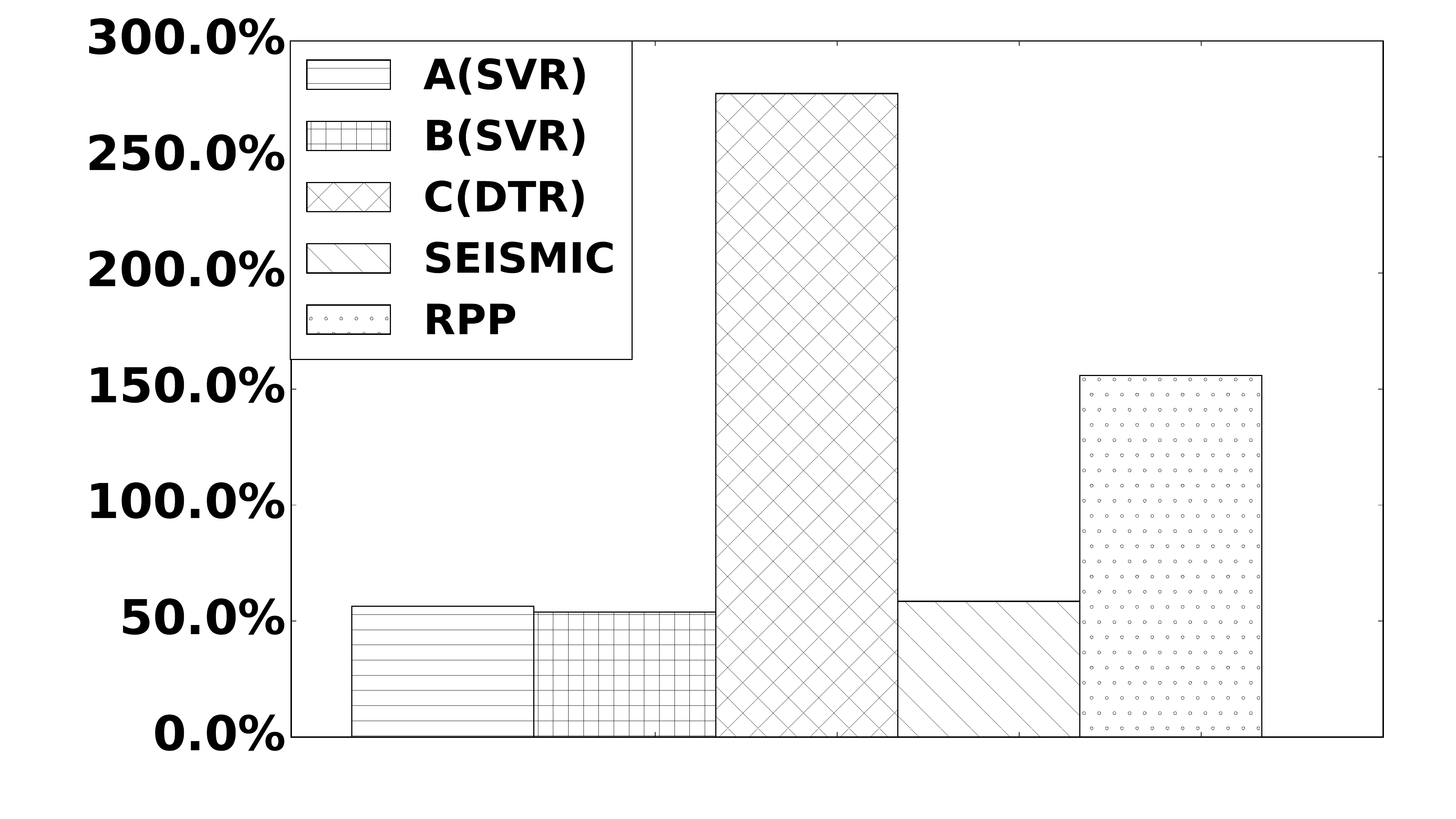}
		\caption{Twitter Dataset: APE}
		\label{fig:reg_ape}
	\end{subfigure}%
	\hfill
	\begin{subfigure}{.24\textwidth}
		\centering
		\includegraphics[height=2.5cm]{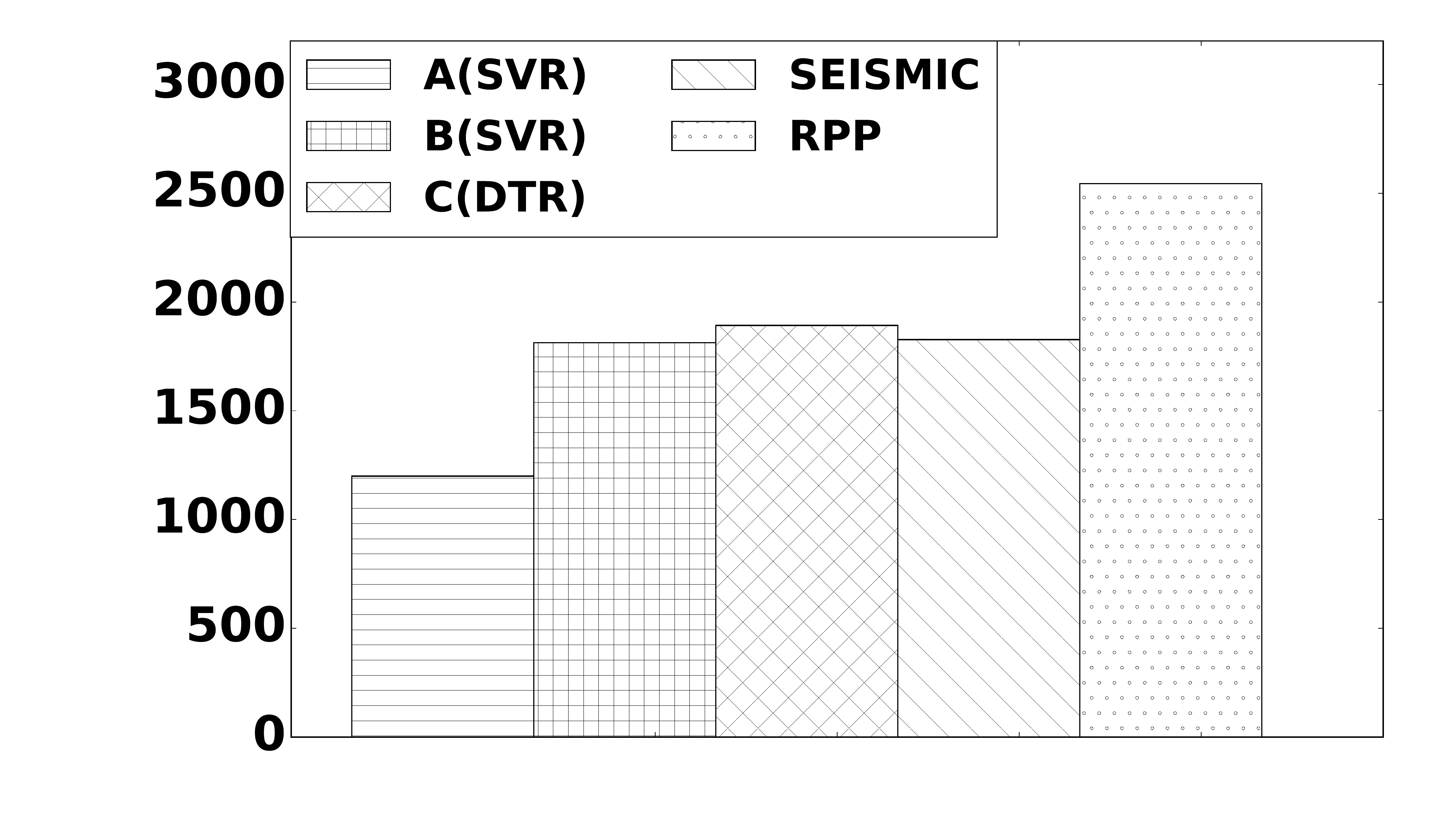}
		\caption{Twitter Dataset: RMSE}
		\label{fig:reg_rmse}
	\end{subfigure}

	\begin{subfigure}{.24\textwidth}
		\centering
		\includegraphics[height=2.5cm]{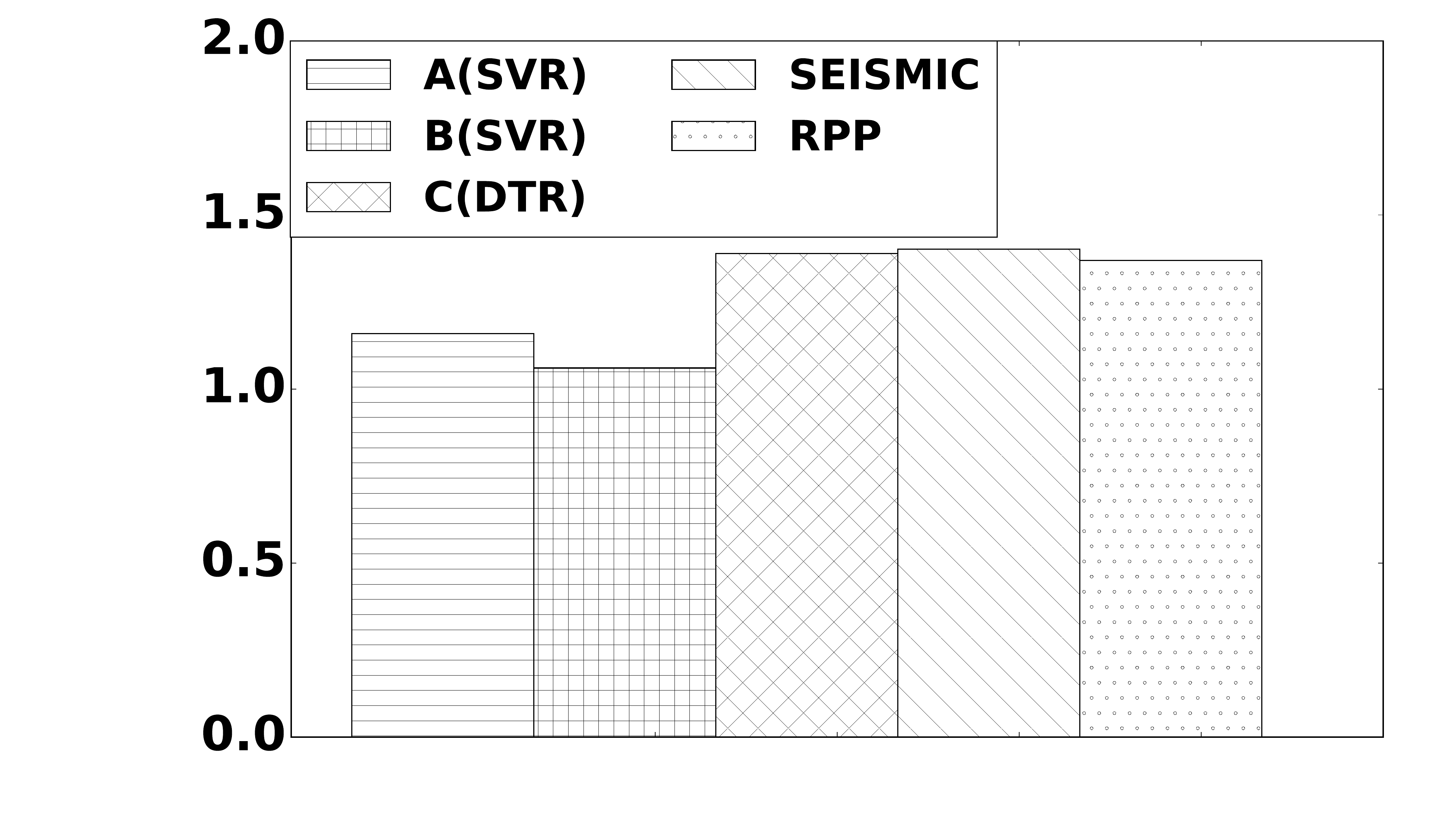}
		\caption{Twitter Dataset: RMSLE}
		\label{fig:reg_rmsle}
	\end{subfigure}
	\hfill
	\begin{subfigure}{.24\textwidth}
		\centering
		\includegraphics[height=2.5cm]{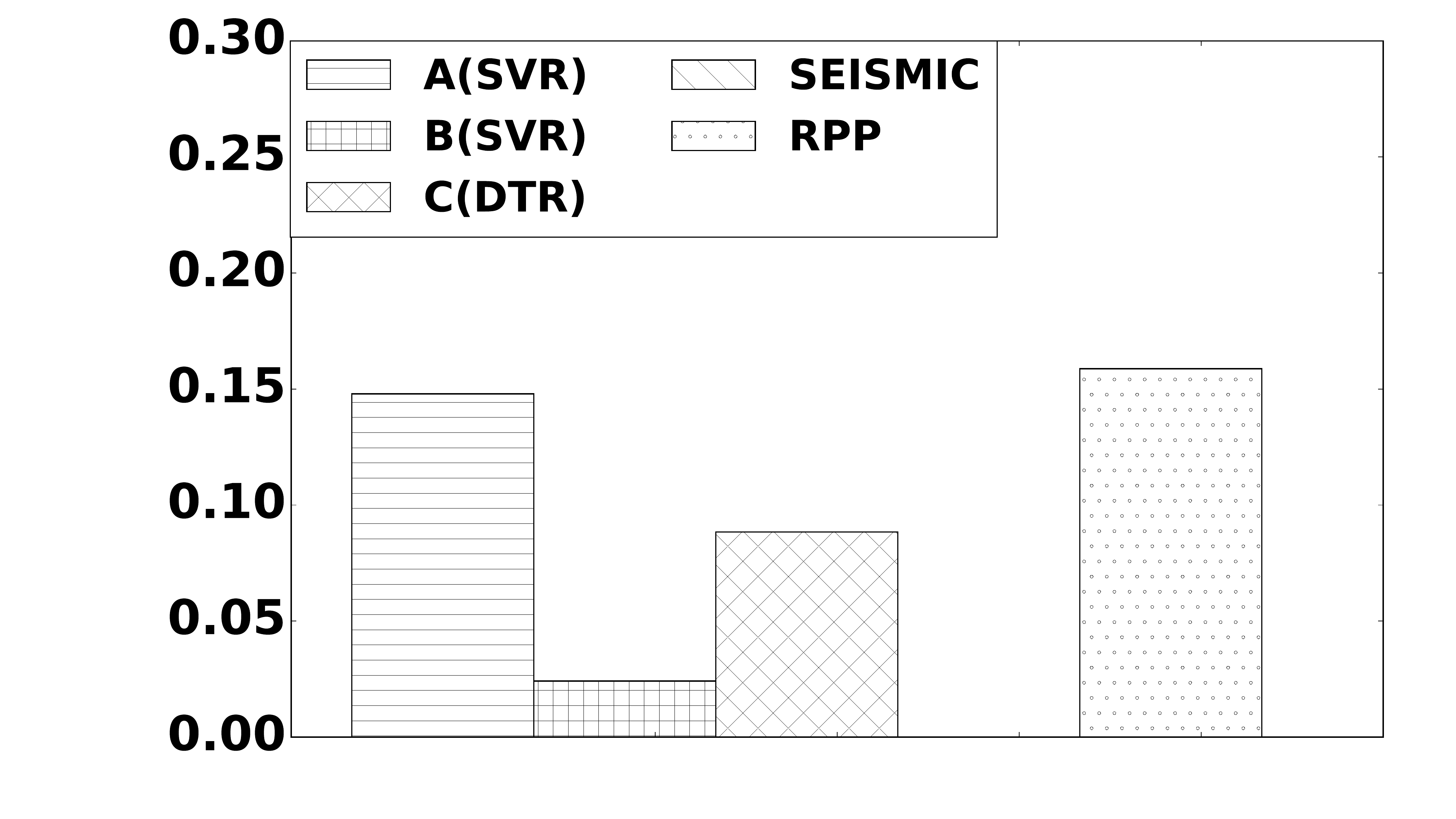}
		\caption{Twitter Dataset: Top 10\% Coverage}
		\label{fig:reg_topcov}
	\end{subfigure}

	\begin{subfigure}{.24\textwidth}
		\centering
		\includegraphics[height=2.5cm]{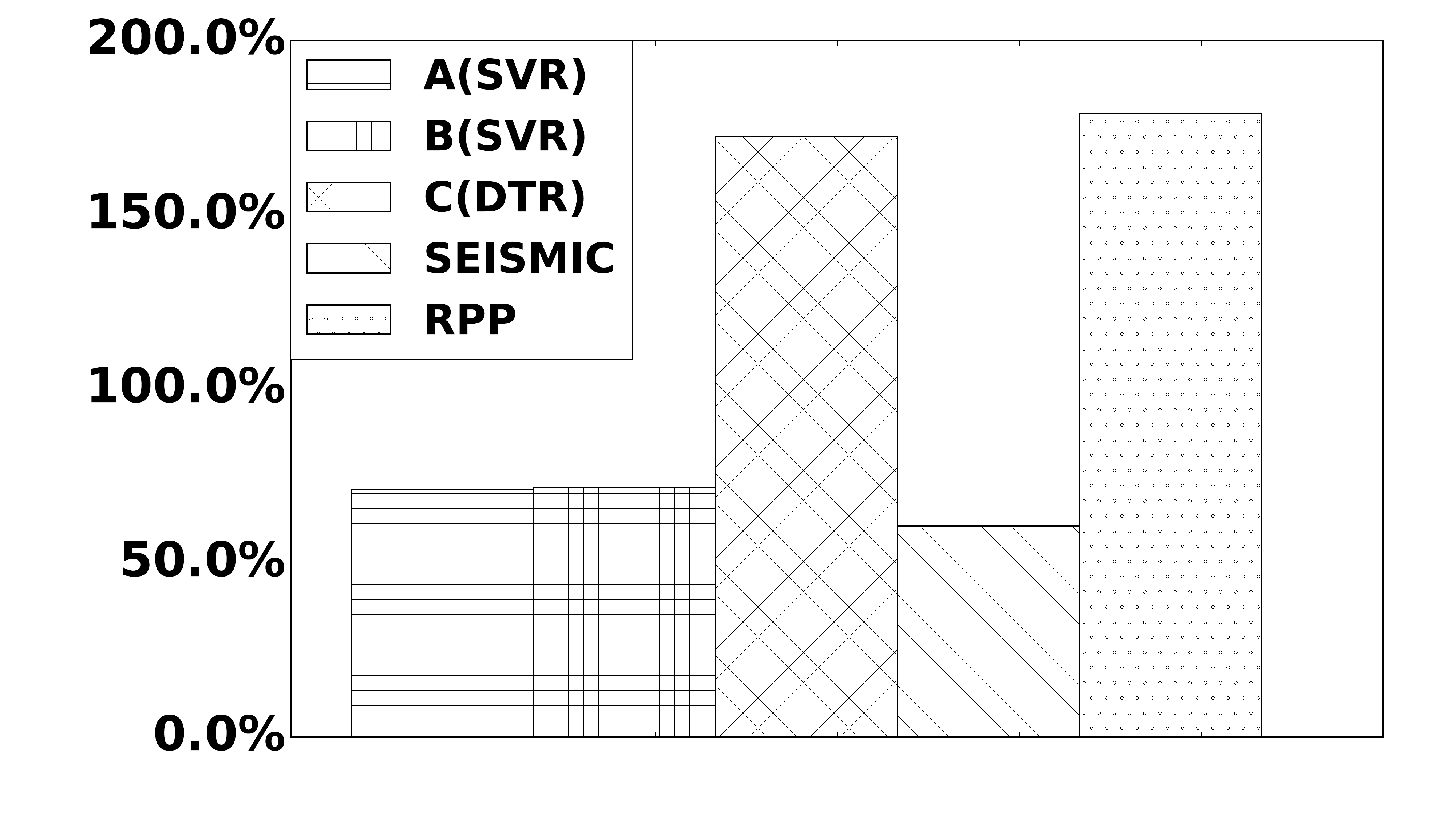}
		\caption{Weibo Dataset: APE}
		\label{fig:reg_ape_w}
	\end{subfigure}%
	\hfill
	\begin{subfigure}{.24\textwidth}
		\centering
		\includegraphics[height=2.5cm]{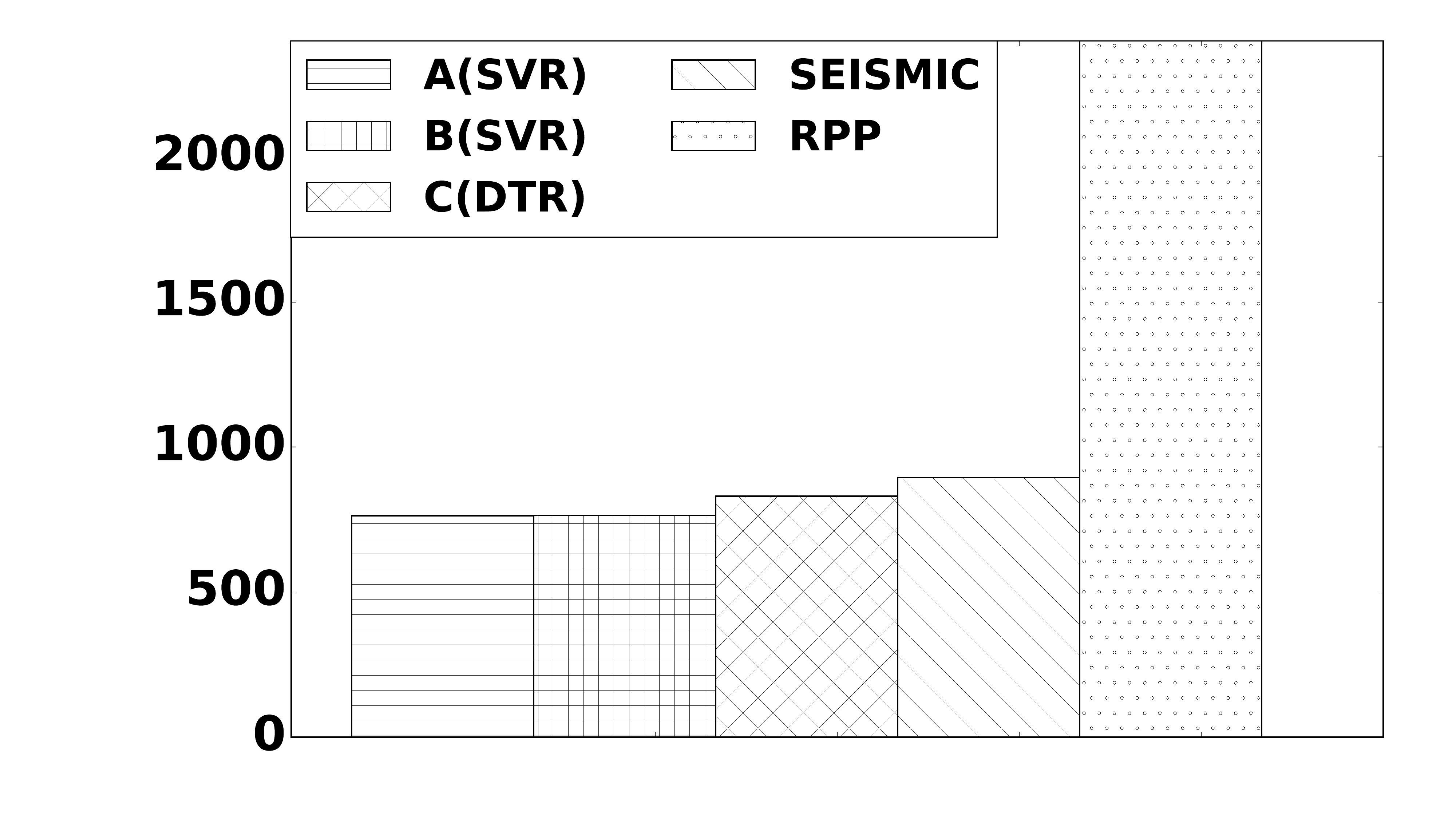}
		\caption{Weibo Dataset: RMSE}
		\label{fig:reg_rmse_w}
	\end{subfigure}

	\begin{subfigure}{.24\textwidth}
		\centering
		\includegraphics[height=2.5cm]{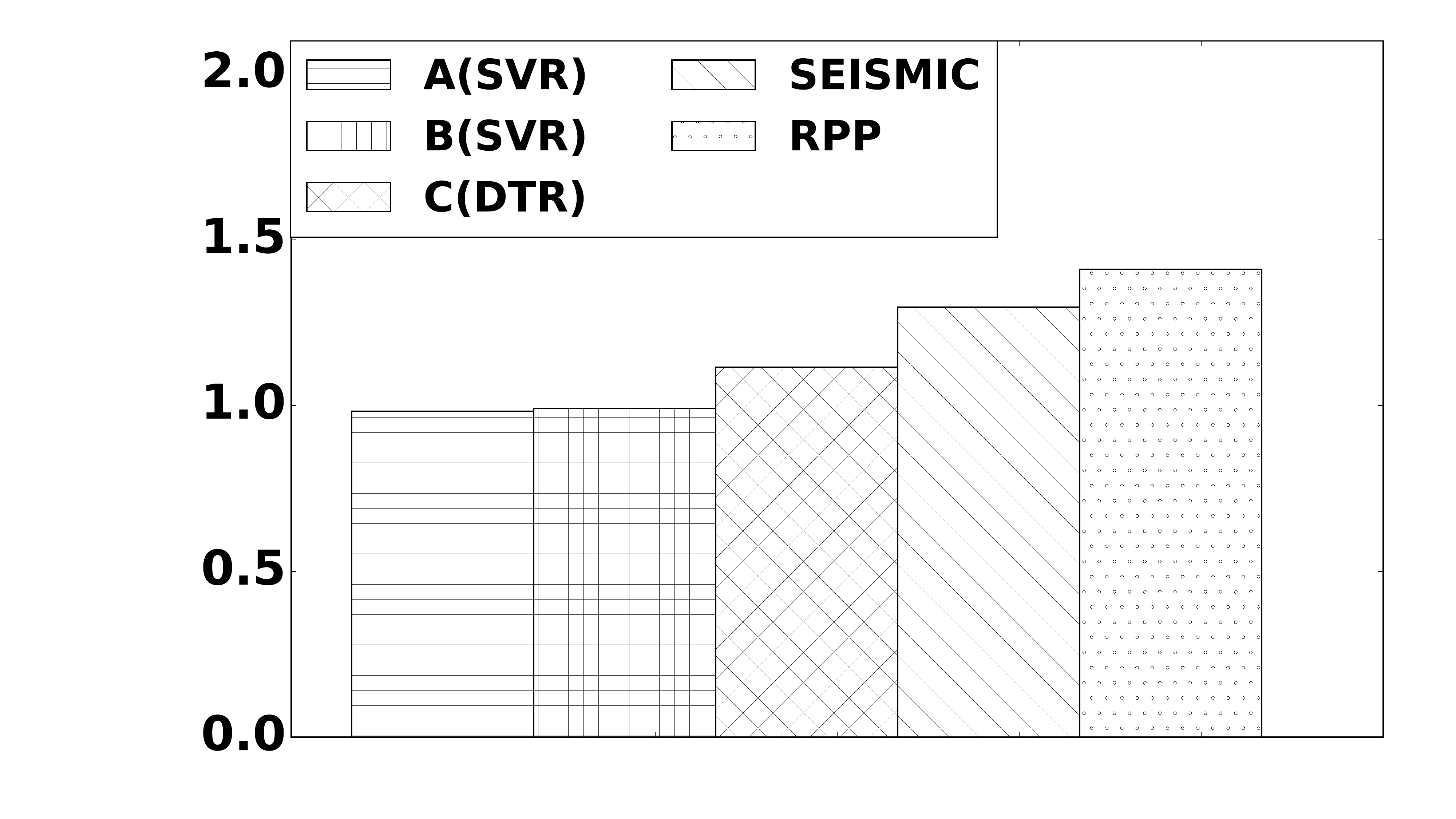}
		\caption{Weibo Dataset: RMSLE}
		\label{fig:reg_rmsle_w}
	\end{subfigure}
	\hfill
	\begin{subfigure}{.24\textwidth}
		\centering
		\includegraphics[height=2.5cm]{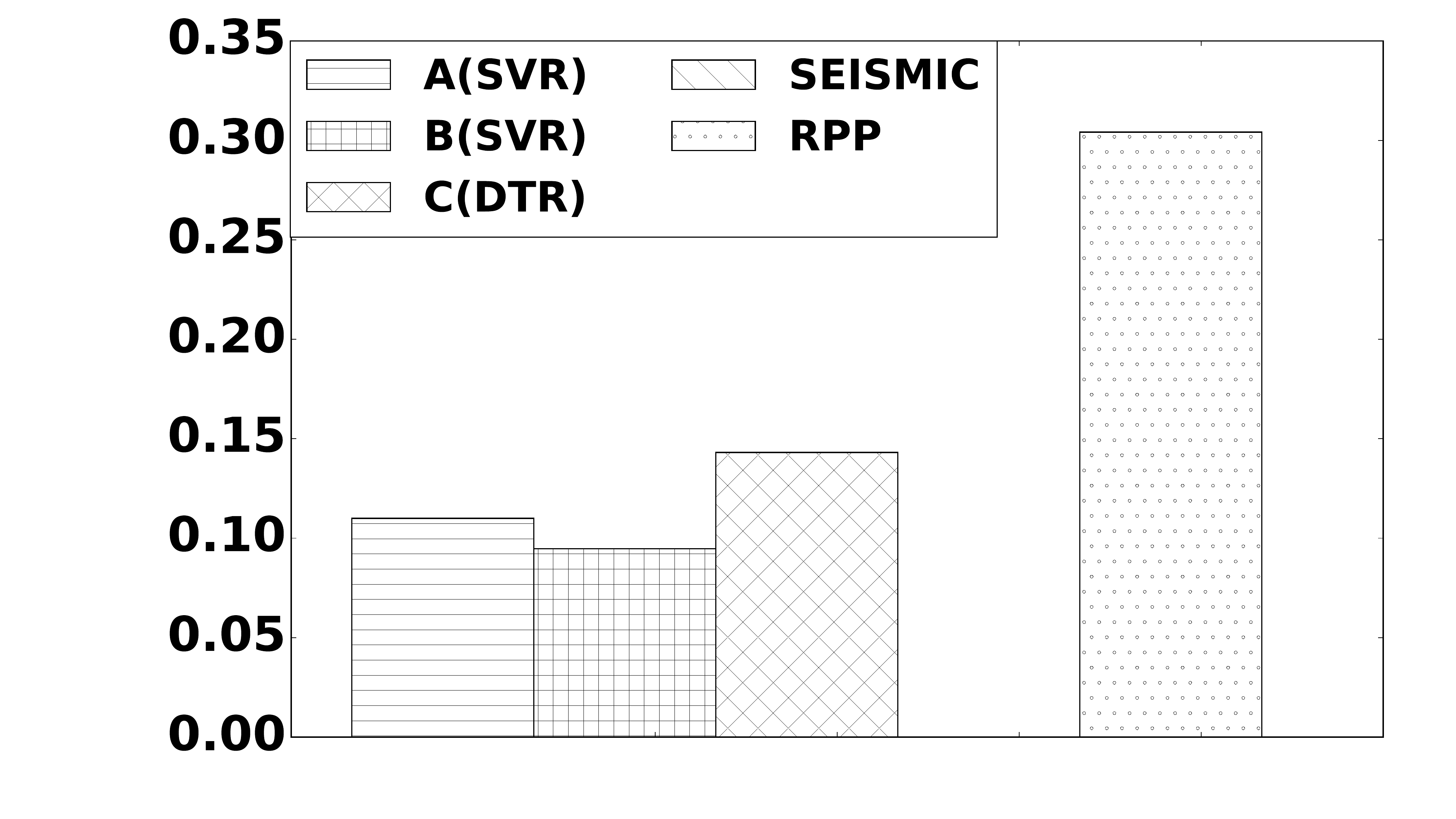}
		\caption{Weibo Dataset: Top 10\% Coverage}
		\label{fig:reg_topcov_w}
	\end{subfigure}
	\caption{Regression results}
	\label{fig:reg_res}
\end{figure}

For the Twitter dataset, Fig.~\ref{fig:reg_ape}, \ref{fig:reg_rmse}, \ref{fig:reg_rmsle} and \ref{fig:reg_topcov} show the experimental results for the regression problem.
Feature based methods and SEISMIC outperform RPP and method C w.r.t. APE.
Concerning RMSE, method A shows more predictive power than others.
As to RMSLE, feature based methods result in less error than the other two categories.
From the aspect of Top 10\% coverage, RPP, method A are more likely to track the trending cascades than others.

Fig.~\ref{fig:reg_ape_w}, \ref{fig:reg_rmse_w}, \ref{fig:reg_rmsle_w} and \ref{fig:reg_topcov_w} show the regression result for the Weibo dataset,
Regarding APE, SEISMIC, method A and B have comparable performance and outperform others.
In terms of RMSE, method A, B are measured to be more predictive than the rest.
Feature based methods also make predictions with least RMSLE.
For top 10\% coverage, RPP is more likely to detect popular cascades than others.

An interesting observation is that the prediction accuracy measured by different error metrics can be contrary to each other.
For example, in Fig.~\ref{fig:reg_ape}, compared to SEISMIC, prediction made by method C results in more error measured by APE, however, comparable error w.r.t. RMSE and less error regarding RMSLE (See Fig.~\ref{fig:reg_rmse} and~\ref{fig:reg_rmsle}).
This implies that it is better for researchers to show more than one type of error for evaluation of regression results.

\subsection{Classification}
We show the precision, recall and F1 score for the viral class with all the three percentile thresholds. For each dataset, we choose the 50th, 75th and 90th percentile of the final size of all cascades as the thresholds for assigning the cascades into viral or non-viral class. The number of samples in each class is shown in Table~\ref{tbclf}. Thus we can evaluate the cascade prediction methods with balanced and imbalanced classes. For each method, we only show the best result among those produced by different classifiers or various training methods. As a result, for feature based methods, random forest outperforms others. 
While for point process based methods we treat cascade size predicted by setting prediction time as $\left\{2,4,6,8,10\right\} \times \mathbf{t_v}(t)[50]$ as features. Here we show the results produced by classifiers trained by these features.

\begin{figure}
	\centering
	\begin{subfigure}{.24\textwidth}
		\centering
		\includegraphics[height=2.5cm]{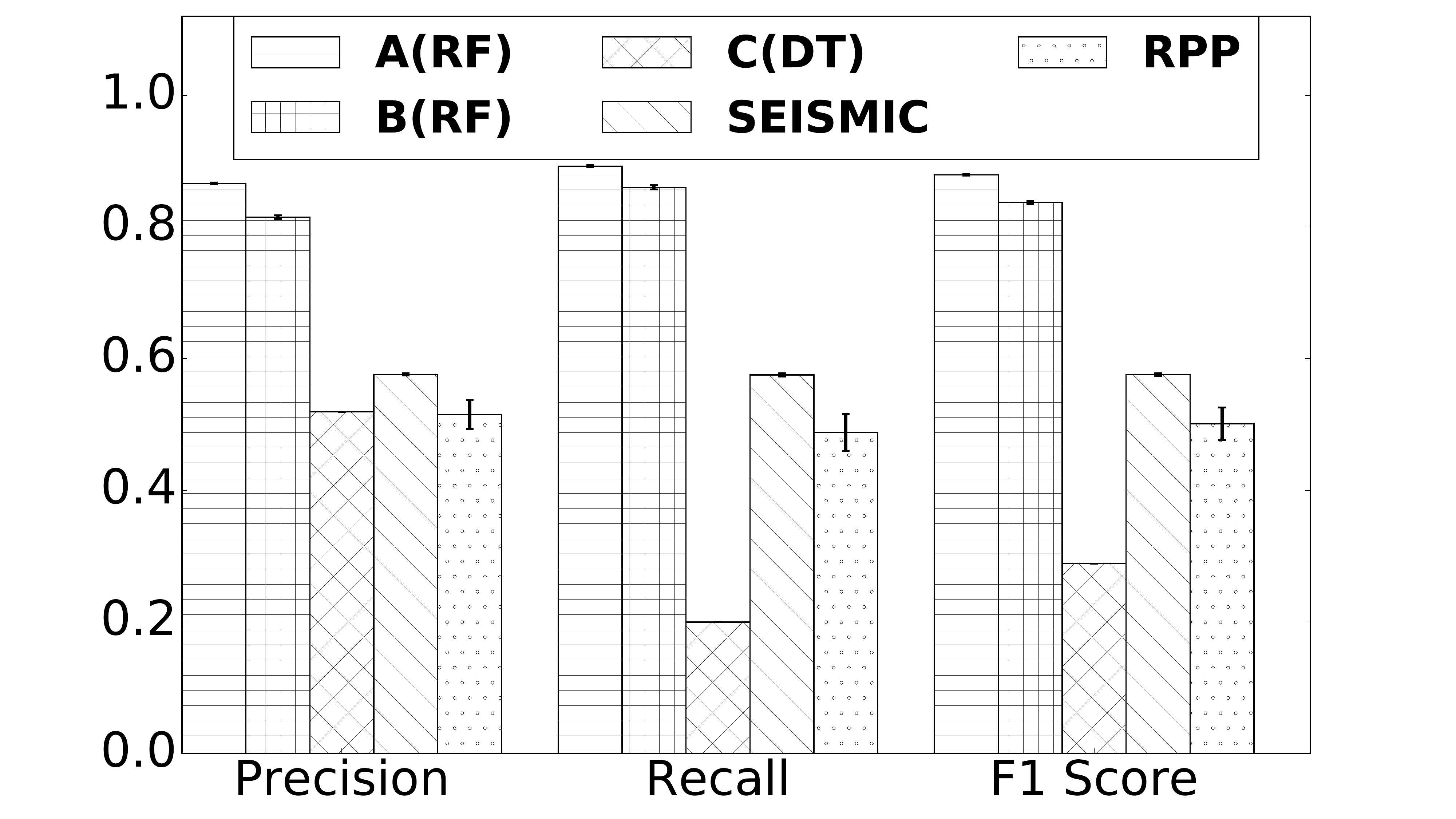}
		\caption{Twitter Dataset: 50th percentile}
		\label{fig:clf_th0}
	\end{subfigure}%
	\hfill
	\begin{subfigure}{.24\textwidth}
		\centering
		\includegraphics[height=2.5cm]{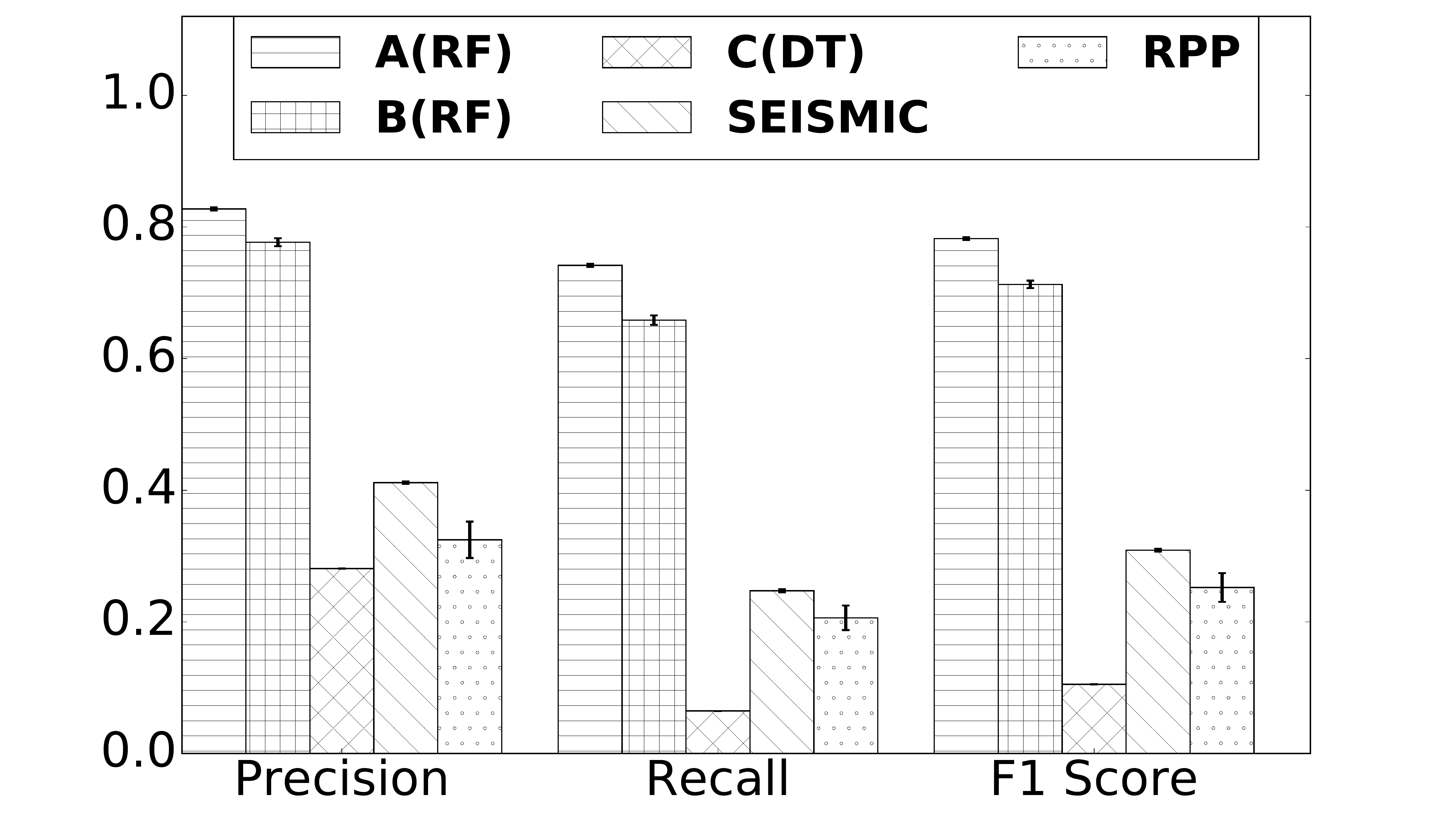}
		\subcaption{Twitter Dataset: 75th percentile}
		\label{fig:clf_th1}
	\end{subfigure}
	\begin{subfigure}{.24\textwidth}
		\centering
		\includegraphics[height=2.5cm]{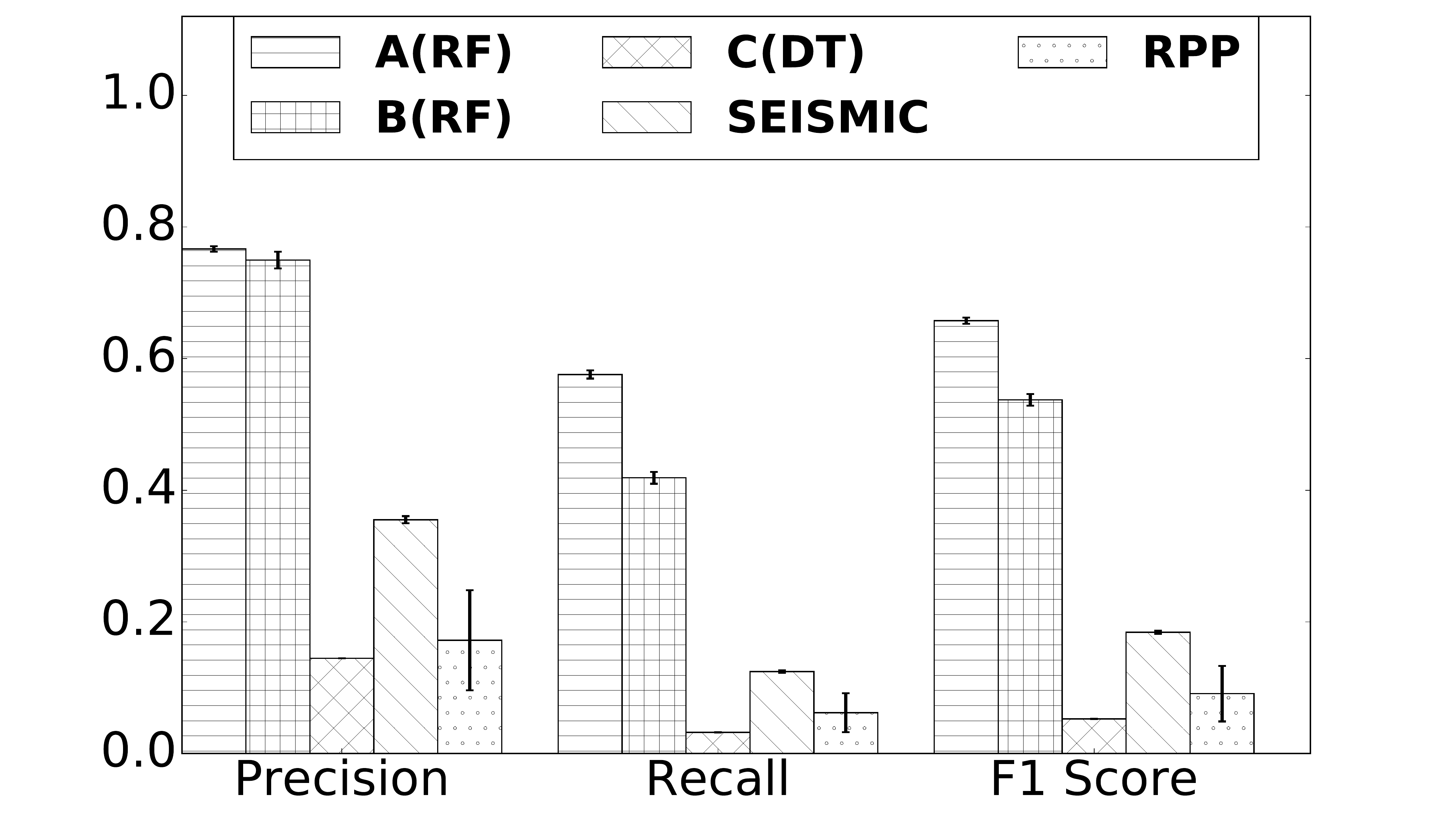}
		\caption{Twitter Dataset: 90th percentile}
		\label{fig:clf_th2}
	\end{subfigure}
	\begin{subfigure}{.24\textwidth}
		\centering
		\includegraphics[height=2.5cm]{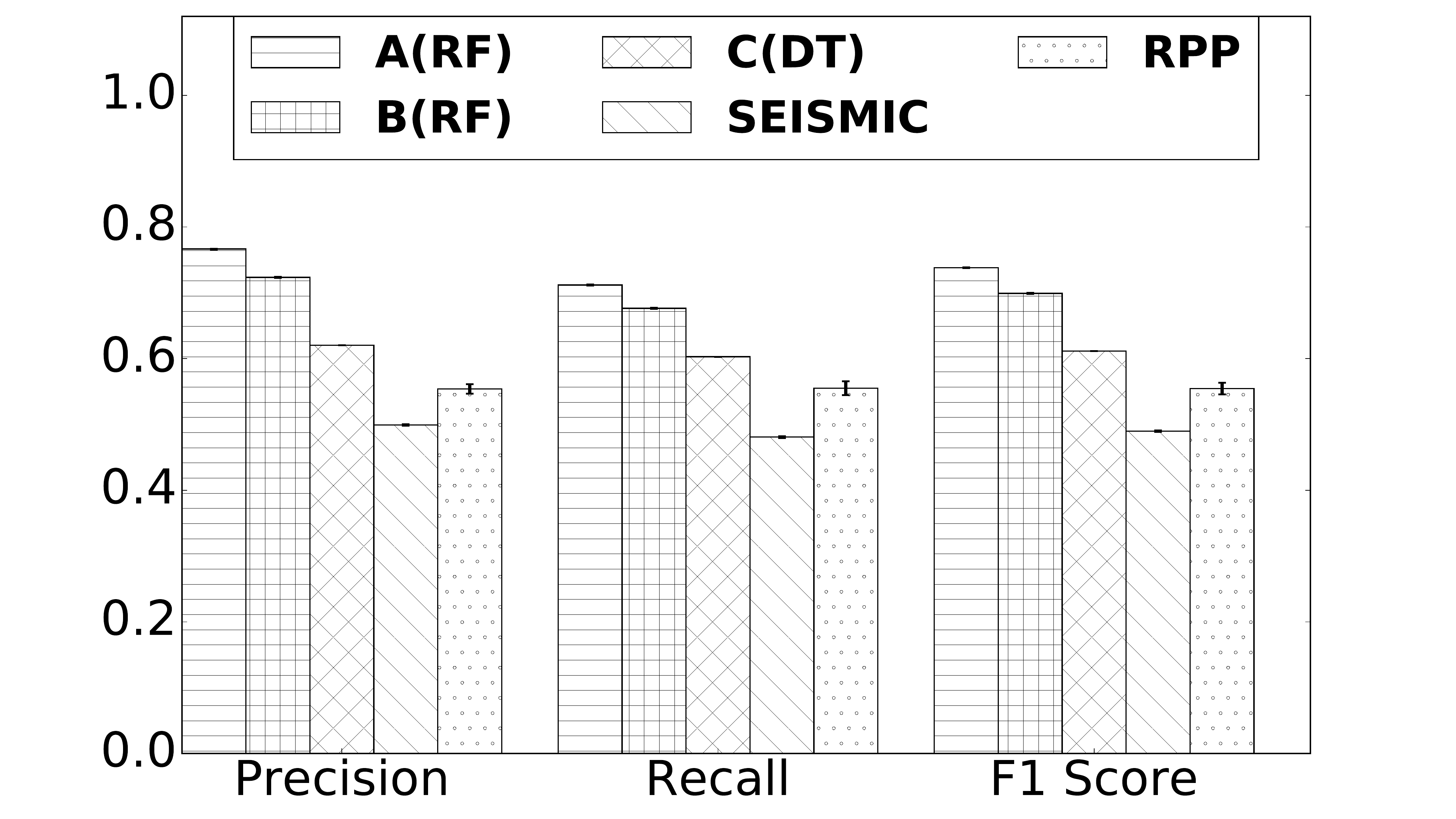}
		\caption{Weibo Dataset: 50th percentile}
		\label{fig:clf_th0_w}
	\end{subfigure}%
	\hfill
	\begin{subfigure}{.24\textwidth}
		\centering
		\includegraphics[height=2.5cm]{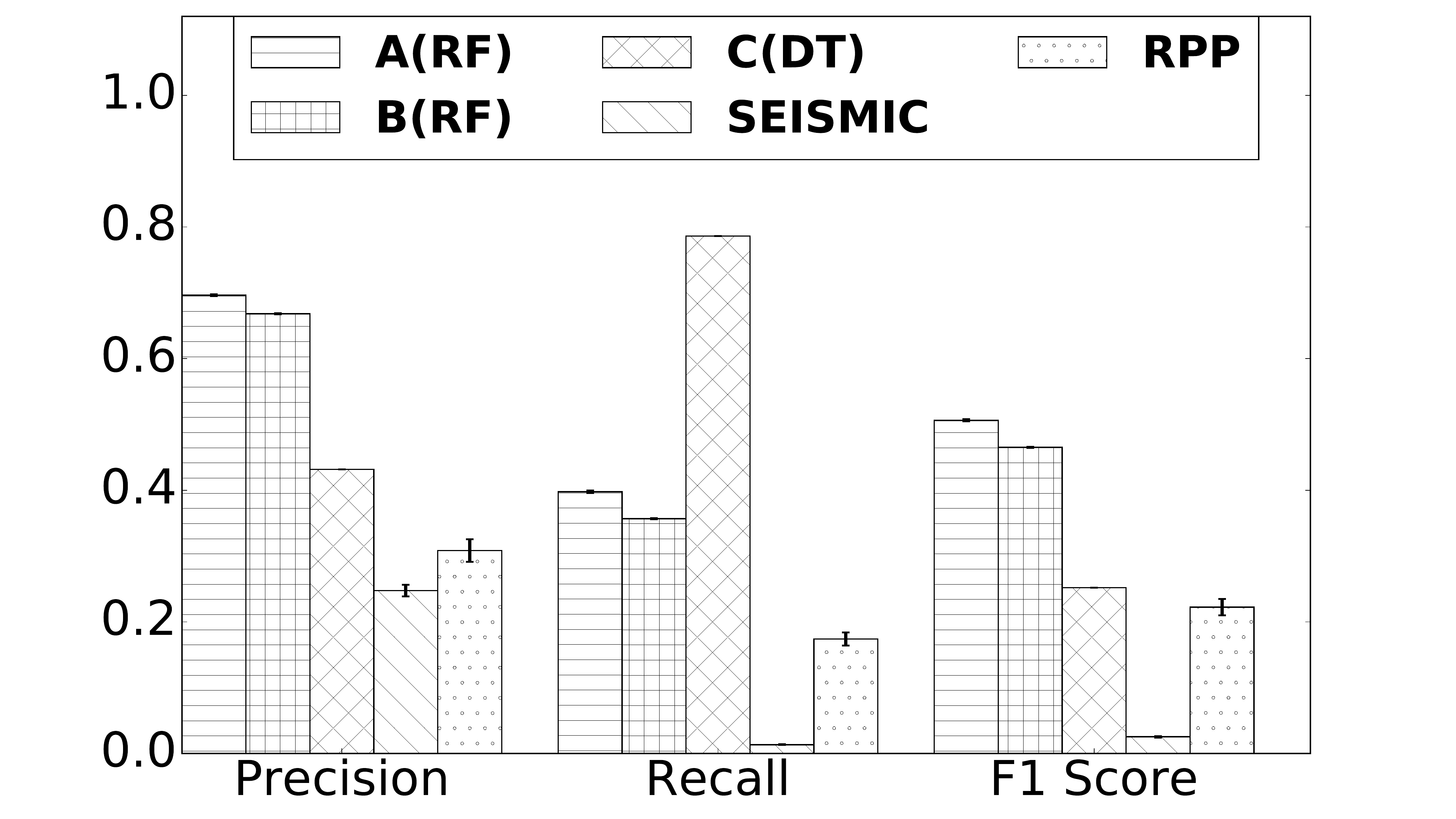}
		\caption{Weibo Dataset: 75th percentile}
		\label{fig:clf_th1_w}
	\end{subfigure}
	\begin{subfigure}{.24\textwidth}
		\centering
		\includegraphics[height=2.5cm]{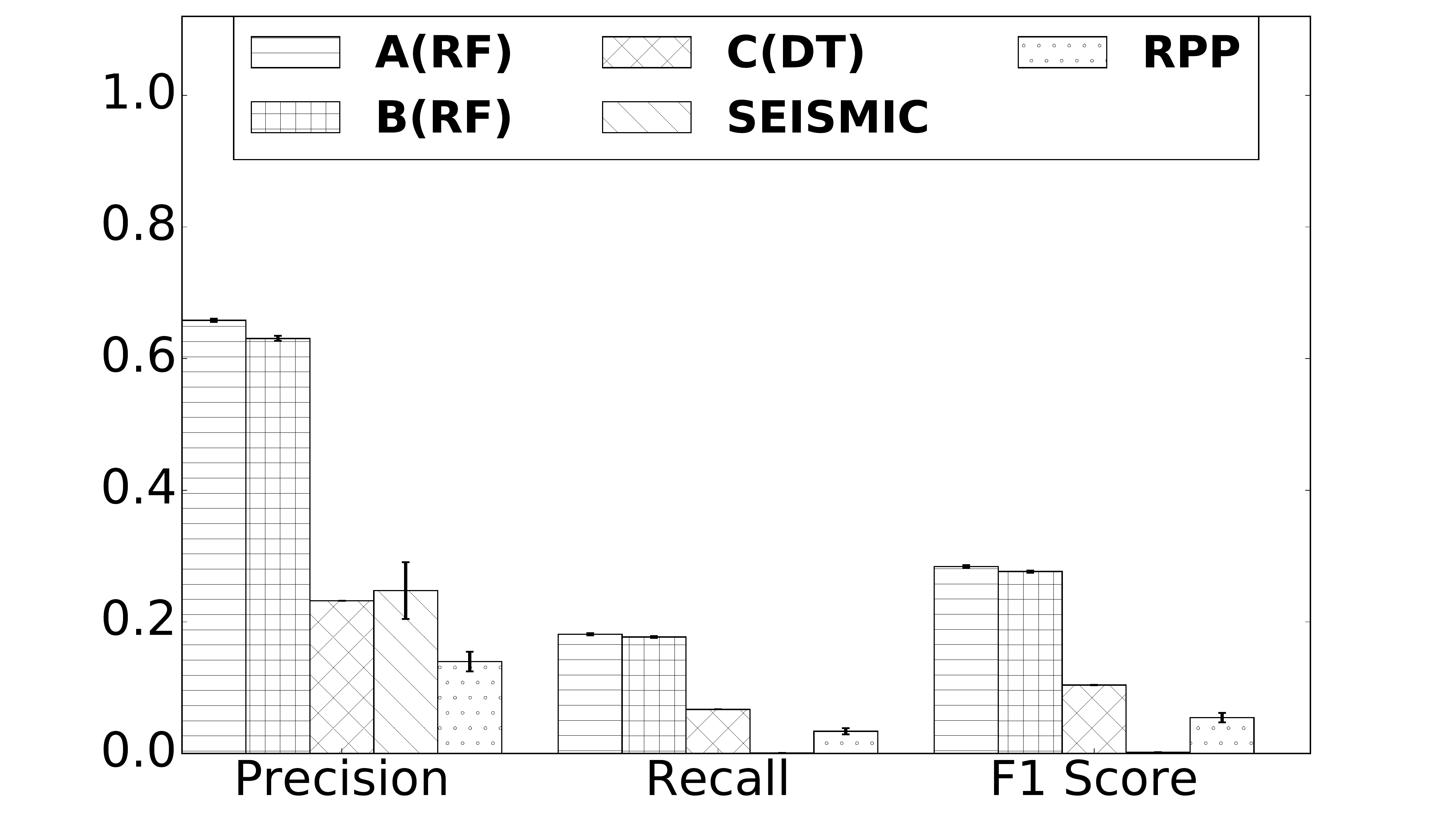}
		\subcaption{Weibo Dataset: 90th percentile}
		\label{fig:clf_th2_w}
	\end{subfigure}
	\caption{Classification results: error bar stands for one standard deviation.}
	\label{fig:clf_res}
\end{figure}

Fig.~\ref{fig:clf_th0}, \ref{fig:clf_th1} and \ref{fig:clf_th2} show the classification results for the Twitter dataset.
With all three thresholds, feature based methods A and B outperform others.
 In addition, they also show more robustness than others to imbalance of two classes in dataset.
In terms of point process based methods, SEISMIC outperforms RPP especially when the two class are imbalanced.
RPP suffers from relatively large standard deviation, as the Newton's method is not always able to achieve convergence. Thus the parameters learned through the MLE approach can vary as a result from random initialization.
Method C (eigenvector centrality) shows little predictive power with any of the three thresholds for the Twitter dataset, even if it outperforms other centrality based methods.

For the Weibo dataset, as shown in Fig.~\ref{fig:clf_th0_w}, \ref{fig:clf_th1_w} and \ref{fig:clf_th2_w}, feature based methods outperform others again with all three thresholds.
Regarding point process based methods, contrary to the results for Twitter dataset, RPP achieves better F1 score than SEISMIC when threshold value becomes large.
Method C (eigenvector centrality) performs comparably to RPP.

\subsection{Run time}
In this subsection, we show the run time of tasks for the cascade prediction methods considered in this paper.
On one hand, preprocessing, computation of centralities and features suffer from high overhead as immense amount of data needs to be loaded.
The run time of these tasks are listed in Table~\ref{tbrtime}.
On the other hand, training and prediction tasks barely have the overhead issue.  

\noindent\textbf{Preprocessing:} We carry out the community detection task by the java implementation of Louvain algorithm~\cite{waltman2013smart} with 10 random start and 10 iterations for each start.
For computation of centralities, we load edgelist of the social networks as a graph object in igraph-python~\cite{igraph}.
As shown in Table~\ref{tbrtime}, community detection, computation of Pagerank and loading graph are the tasks suffer the most when the size of dataset increases. Community detection, computation of Pagerank and loading graph for the Weibo dataset take 80.32, 66.855 and 19.80 times the run time of those for the Twitter dataset respectively.

\noindent\textbf{Computation of Features:}
As shown in Table~\ref{tbrtime}, for the feature computation task, it takes method B 12.37 and 8 times the run time method of A for the Twitter Dataset and the Weibo Dataset respectively. To explain this observation, an analysis of what computation is carried out in each iteration for method A and B.
For method A, computation of the features can be done without loading the graph (a heavy overhead). Moreover, for each cascade, method B also requires expensive computation of shortest path length for each pair of nodes in cascade subgraphs and size of 2-hop neighborhood.

\noindent\textbf{Training and Prediction:}
The run time of training and prediction is not directly related to the size of the social network.
On one hand, it is correlated to the number of cascades for training and prediction. On the other hand, it is decided by the complexity of the method: for example, number of parameters to be learned, the complexity for learning each paramter and the comsumption to work out the prediction.
Here we only measure the run time for solving the classification problem.
We run each method with single process, overhead run time such as graph loading is ignored.
For feature based methods the training and prediction time are also correlated to the number of features.
For centrality based methods, we only show the run time for k-shell (method C) as all methods in this category are trained and tested with one feature: the centrality measure of the root node.
Compared to RPP, SEISMIC is a deterministic method with closed form solution. The run time for each sample can be distributed with little variance.
For the RPP method, as the log-likelihood function is non-convex, it is not guaranteed that global maximum can be reached in limited number of iterations. Therefore, the run time for a sample running out of the maximum number of iterations can be thousands times that of another, which reaches the convergence condition in the first iteration.
As the log-likelihood function of RPP is twice-differentiable, Newton's method can be applied.
In our experiments, with the maximum number of iterations setted as 100, the convergence is more likely to be achieved by Newton's method than gradient descent.
Thus we only show the run time of RPP with Newton's method.

\begin{figure}
	\centering
	\begin{subfigure}{.24\textwidth}
		\centering
		\includegraphics[height=2.5cm]{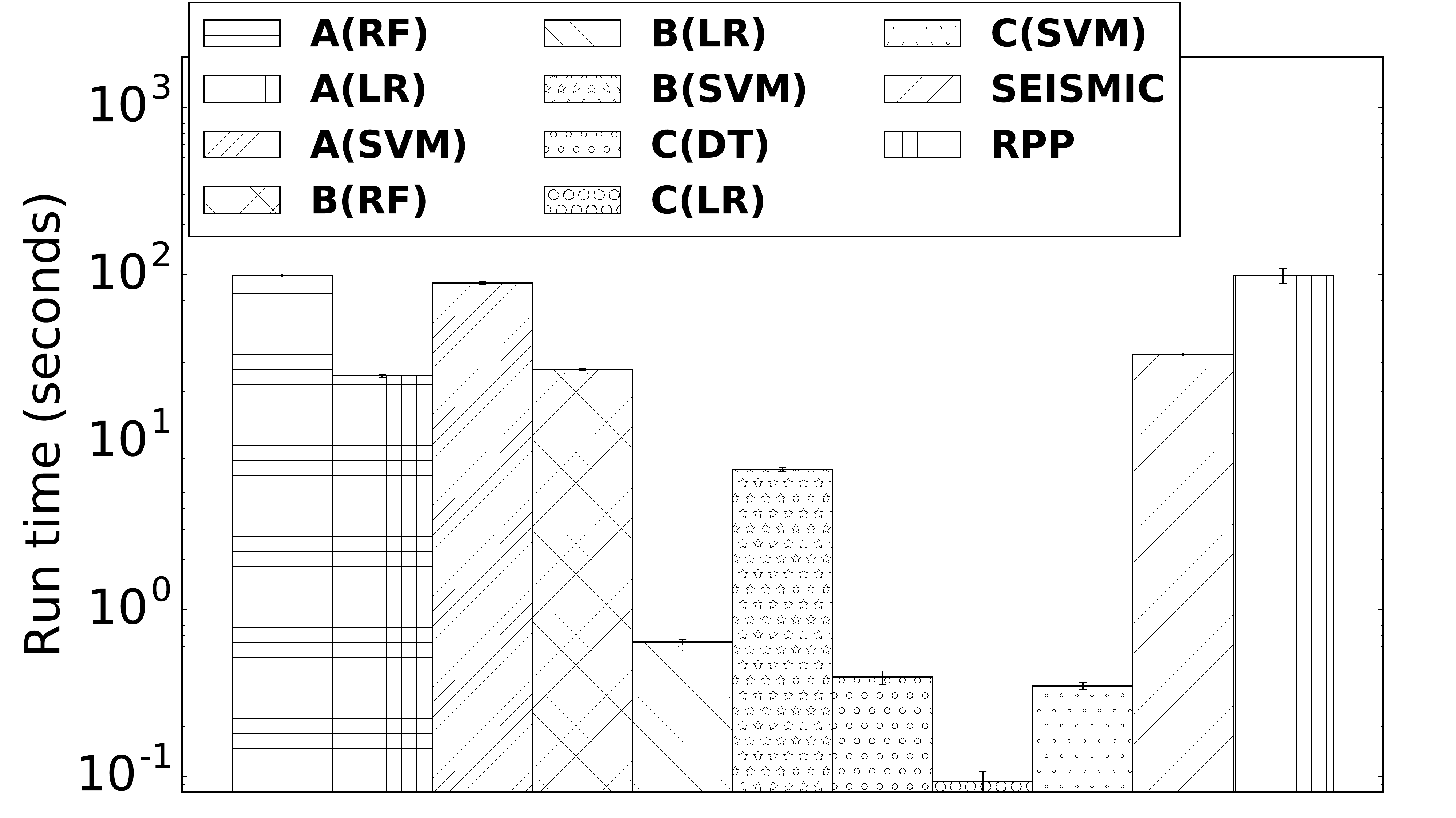}
		\caption{Twitter Dataset}
		\label{fig:run_time}
	\end{subfigure}%
	\hfill
	\begin{subfigure}{.24\textwidth}
		\centering
		\includegraphics[height=2.5cm]{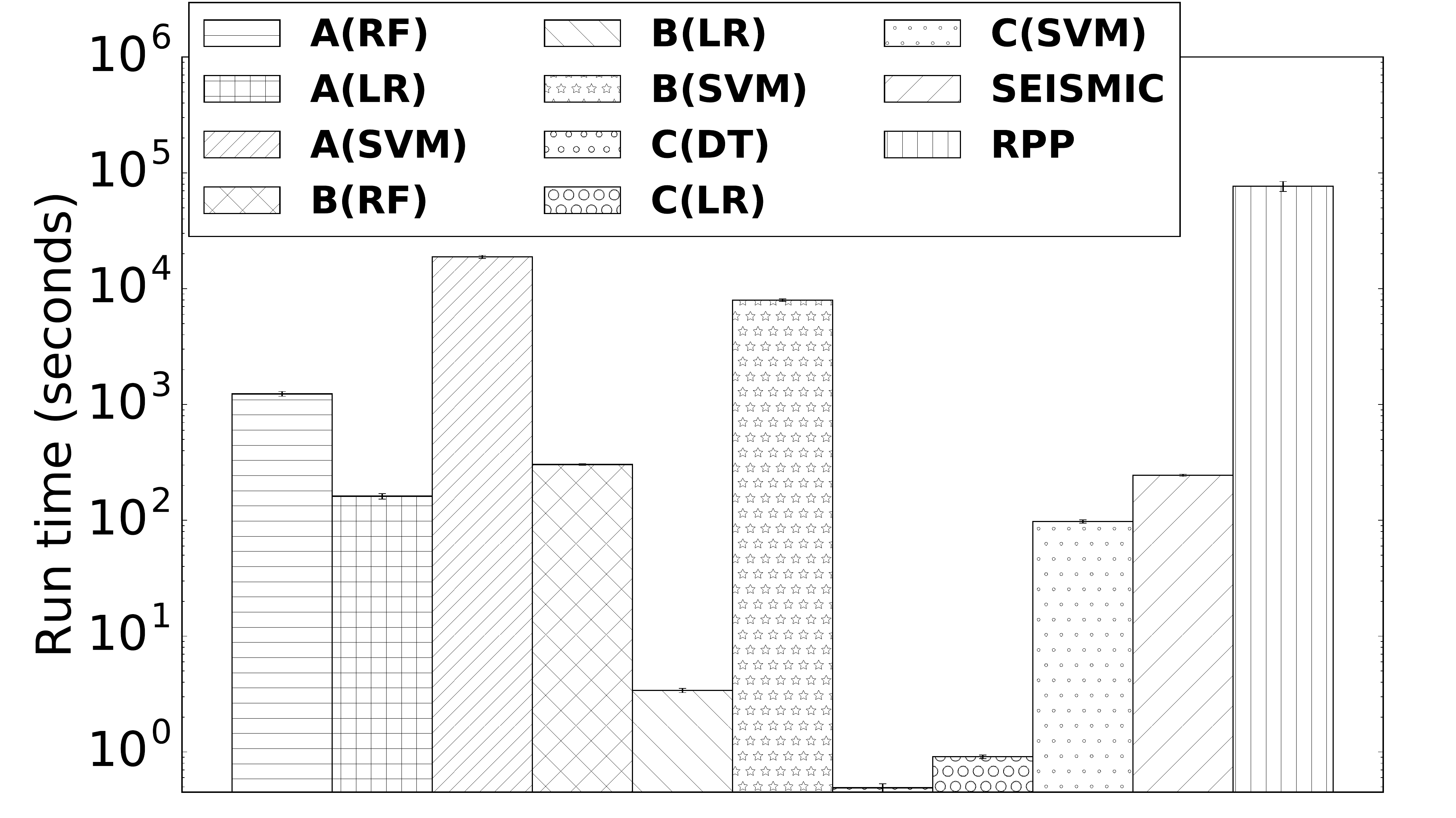}
		\caption{Weibo Dataset}
		\label{fig:run_time_weibo}
	\end{subfigure}
	\caption{Run Time of Trainig and Prediction}
	\label{fig:rt}
\end{figure}

Fig.~\ref{fig:rt} shows the run time for each method to complete training and prediction tasks for all cascades in the two datasets.
For feature based methods, it shows the run time needed for random forest (RF), SVM and logistic regression (LR).
For method C, it shows that of decision tree (DT), SVM and logistic regression (LR).

 Concerning the Twitter dataset (See Fig.~\ref{fig:run_time}), taking advantage of decent implementation of classifiers, feature based methods comparable run time to point process based methods w.r.t. the training and prediction task with random forest and SVM (rbf kernel).
 
 For the Weibo dataset, as shown in Fig.~\ref{fig:run_time_weibo}, the run time feature based methods comsume is comparable to SEISMIC with random forest. But the SVM with rbf kernel suffers from the order-of-magnitude increase of the number of training and testing samples. Thus leads to the observation that the run time becomes approximately 10 times that of random forest.
 
 Comparing Fig.~\ref{fig:run_time_weibo} with Fig.~\ref{fig:run_time}, the run time of RPP method increases the most. 
 This means it is much more difficult for the Newton's method to converge for samples in the Weibo datasets. There are two possible reasons to explain this: 1). the uniform distribution used in random initialization can not produce good initial values that are closed to local optimal points; 2). the choice of log-norm distribution as function $f_d(t;\theta_d)$ can not provide fairly good description of cascades in this dataset.

\begin{table}
	\caption{Run time: Preprocessing \& Feature Computation}
	\centering
	\renewcommand{\arraystretch}{1.2}
	\begin{tabular}{ | p{2cm}|p{1.75cm}|p{1.5cm} |p{1.5cm}| } 
		\hline
		Type & Task & Total time (s) & Time per sample (s)\\
		\hline
		\multicolumn{4}{|c|}{Twitter Dataset} \\ 
		\hline
		\multirow{6}{*}{Preprocessing}& Louvain & 275 & -- \\
		& Loading Graph & 60.033 & -- \\
		& Degree & 0.016 & --\\
		& K-shell & 2.757 & -- \\
		&Eigenvector & 20.444 & -- \\
		&Pagerank & 26.298 & -- \\
		\hline
		
		\multirow{2}{2cm}{Feature Computation} & A & 267.144 & 0.018\\
		
		& B & 3252.7562&0.2227 \\
		\hline
	
		\multicolumn{4}{|c|}{Weibo Dataset}\\ 
		\hline
		\multirow{6}{*}{Preprocessing}& Louvain & 22087 & -- \\
		& Loading Graph & 1188.486 & -- \\
		& Degree & 0.045 & --\\
		& K-shell & 139.128 & -- \\
		&Eigenvector & 391.140 & -- \\
		&Pagerank & 1758.164 & -- \\
		\hline
		\multirow{2}{2cm}{Feature Computation} & A & 11181.453 & 0.110 \\
		& B& 87651.213 & 0.883\\
		\hline
		
	\end{tabular}
	\label{tbrtime}
\end{table} 

\section{Related Work}
\label{sec:related}
\noindent\textbf{Influence Maximization}
Since the proposal of Influence maximization problem by Kempe et al.~\cite{kempe2003maximizing}, related work emerged, focusing on estimation of influence for a selected set of nodes that can be measured by expected number of infectees under a certain influence model, such as~\cite{chen2009efficient} and~\cite{goyal2011celf++}.
Recently, a scalable randomized algorithm designed by Du et al.~\cite{du2013scalable} estimates influence initiated by selected source nodes and thus select seed set with maximum expected influence.

\noindent\textbf{Cascade Prediction}
Although in~\cite{pei2014searching}, k-shell and heuristics of k-shell were shown to be effective indicator of long-term influence of nodes, in~\cite{shakarian2015diffusion}, experimental results showed that the shell number of the root node is not effectively predictive in the cascade by cascade scenario.
Feature based methods from Jenders et al.~\cite{jenders2013analyzing} Chen et al.~\cite{cheng2014can} were designed to solve the cascade prediction problem formulated as binary classification on balanced dataset, however, these methods are more or less dependent on content features from specific social media sites.
Ma et al.~\cite{ma2013predicting} focused on applying content features to classify hashtag cascades by how much their size increases.
Regarding to point process based methods, model designed with the intuition of mutual exciting nature of social influence, Zhou et al~\cite{zhou2013learning} applied multi-dimensional Hawkes process to rank cascades (memes) by their popularity.
Recently, the model introducted by Yu et al.~\cite{yu2015micro} combined feature engineering and human reaction time distribution function widely used in point process based methods to aggregate adoptions in \textit{subcascades} for cascade prediction.
Besides feature based methods and point process based methods studied in this paper, knowledge from related research fields could also be applied to cascade prediction. Goyal et al.~\cite{goyal2011data} proposed the credit distribution model to learn pair-wise influence based on IC model proposed by Kempe et al.~\cite{kempe2003maximizing}.
Cui et al.~\cite{cui2013cascading} proposed a feature selection approach for binary classification of cascades.
Wang et al.~\cite{wang2015learning} proposed a model to decouple the influence measured in a pair-wise way
 into two latent vectors representing influence and susceptibility of a node.
 This work differs from all the past efforts in that it is the most thorough comparison of methods general enough to be applied to different datasets without relying on features specific to a certain social media site.

\section{Discussion and Conclusion}
\label{sec:conclusion}
In this paper, we evaluate three categories of recently proposed methods with both the classification and regression formulaton of cascade prediction.
Feature based methods generally provide better prediction accuracy for the cascade prediction problem, no matter it is considered as classification or regression.
However, they suffer from heavy overhead such as community detection and computation of features.
Random point process based methods enable us to achieve the prediction with little preprocessing but are shown to be less accurate than feature based methods.
The run time of methods in this category can also suffer from the situation when the data can not be well modelled by the proposed density function $\lambda(t)$.

In regression experiments, we find the inconsitancy between evaluation with different error metrics. A method that performs well w.r.t. one metric could result in large error measured by another.
A predictive method should be able to perform fairly well measured by various error metrics.

How to deal with changes in the social network and progress of cascades to update features is the biggest issue that both centrality based and feature based methods encounter.
The heavy overhead introduced by preprocessing and computation of features limits these methods from near real-time prediction.

Point process based methods require little preprocessing and the training and prediction process are parallelable as they consider each cascade is indenpendent of others. This advantage in terms of run time over feature based methods can also be amplified as the size of the social network and the number of cascades. Moreover, point process based methods encounter little cold start problem. These two characteristics of point process based methods make them more suitable for real-time cascade prediction task. But how to secure the accuracy of prediction is the biggest issue for them.
The point process based models are faced with two more problems: sensitivity to scale of time unit and requirement of prediction time as an input variable.
In real-world application, given a early stage cascade, estimation of when it will stop progressing is a non-trivial problem.

On balance, this paper explored various methods in the academic literature of predicting viral information cascades in a more comprehensive manner.
Our aim is to provide important insights into which methods based on graph topology or temporal dynamics performed best - as these results can generalize to a variety of application domains.
In our ongoing work on developing a deplyable system for identifying viral extremist messages, this represents an important consideration.
Our next step is to consider microblog content as well - which tends to be more domain specific.

\section*{Acknowledgments}
Some of the authors are supported through the AFOSR
Young Investigator Program (YIP) grant FA9550-15-1-0159,
ARO grant W911NF-15-1-0282, the DoD Minerva program
and the EU RISE program.

\bibliographystyle{IEEEtran}
\bibliography{bib}
\end{document}